\documentclass[aps,prb,twocolumn,a4paper,10pt]{revtex4-2}
\usepackage{amssymb}
\usepackage{amsmath}
\usepackage{graphicx}
\usepackage{color}
\usepackage{setspace}
\usepackage{tabularx}
\usepackage{makecell}
\usepackage{multirow}
\usepackage{tabularray}
\usepackage{adjustbox}
\usepackage{xurl}
\DeclareMathOperator{\sgn}{sgn}

\begin{document}

\title{An efficient formalism for inertial spin waves: Dzyaloshinskii-Moriya antiferromagnets as case studies}

\date{\today}

\author{De-Yun Zhao$^1$, Ri-Xing Wang$^2$, Meng-Qiu Cai$^1$, Mikhail Cherkasskii$^3$, Peng-Bin He$^1$\footnote[0$^{\dag}$]{$\dag$ hepengbin@hnu.edu.cn}}

\affiliation{$^1$School of Physics and Electronics, Hunan University, Changsha 410082, China \\
$^2$ College of Computer and Electrical Engineering, Hunan University of Arts and Science, Changde 415000, China \\
$^3$Institute for Theoretical Solid State Physics, RWTH Aachen University, DE-52074 Aachen, Germany}

\begin{abstract}
Magnetic inertia, emerging in the ultrafast regime, supports inertial spin waves (SWs) as novel magnetic excitations. Despite considerable efforts devoted to inertial SWs, a systematic formalism for fully characterizing their intrinsic properties, especially chirality and polarization, is still lacking, and inertial SWs in spatially nonuniform magnetic configurations remain poorly explored. Here, we develop a framework for calculating inertial SWs and establish a general definition of their chirality and polarization via the ellipticity angle, a unified parameter encoding frequency sign, phase difference, and elliptical axis ratio. Using this method, we systematically investigate precessional and nutational SWs in uniaxial antiferromagnets with staggered and homogeneous Dzyaloshinskii-Moriya interactions (DMIs), covering uniform collinear, canted, and spiral magnetic configurations. The results reveal that small staggered DMI preserves spin-wave degeneracy, whereas small homogeneous DMI lifts it. Further space-time inversion symmetry breaking in canted and spiral structures fully removes spin-wave degeneracy across the entire Brillouin zone. Long-wavelength nutational SWs behave as backward waves, and flat bands emerge in canted and spiral configurations near a critical inertial relaxation time. In canted and spiral configurations, nutational modes are always lefthanded whereas precessional modes are always righthanded; additionally, the dispersion spectra of the canted configuration can be derived from those of the spiral configuration via band folding. Polarization is wavenumber insensitive for uniform configurations but becomes strongly dispersive for nonuniform ones. This work advances the fundamental understanding of magnetic inertial dynamics and provides theoretical insights for the development of ultrafast magnonic devices.
\end{abstract}

\maketitle

\section{Introduction}

\setlength{\parskip}{0pt}

Magnetic inertia \cite{Ritwik Mondal JMMM}, emerging in the context of ultrafast spin dynamics, has been experimentally confirmed via studies of magneto-inertial resonant excitations \cite{Kumar Neeraj,Vivek Unikandanunni,Anulekha De}. Theoretically, such magnetic inertia is incorporated into the Landau-Lifshitz-Gilbert (LLG) equation as a second-order time derivative term of magnetization. Several mechanisms have been proposed to explain the inertial term in the inertial LLG equation. Within the framework of mesoscopic non-equilibrium thermodynamics, magnetic inertia is obtained by extending the phase space of the magnetization to the degrees of freedom of the angular momentum \cite{M.-C. Ciornei}. Grounded in basic mechanical and electrodynamic principles, magnetic inertia emerges naturally from the fundamental proportionality between angular momentum and magnetization. \cite{J.-E. Wegrowe}. Alternatively, a purely mechanical microscopic model (a circular current loop) can also yield the same inertial terms without presupposing this proportionality \cite{Stefano Giordano}. When memory effects arising from the noninstantaneous interaction between the magnetic system and its environment, such as electron and phonon baths, are taken into account on ultrafast time scales \cite{Harry Suhl,Manfred Fahnle,Satadeep Bhattacharjee,Toru Kikuchi,Utkarsh Bajpai,Pascal Thibaudeau,J. Anders,Fuming Xu,Mario Gaspar Quarenta,Christian Svingen Johnsen}, the inertial term can be traced to the second-order contribution, in time or frequency, of the delayed environmental response. This contribution can be captured, for example, by an expansion of the memory kernel \cite{Harry Suhl,Manfred Fahnle,Satadeep Bhattacharjee,Toru Kikuchi,Pascal Thibaudeau,J. Anders}, a low-frequency expansion of the self-energy \cite{Utkarsh Bajpai,Fuming Xu,Christian Svingen Johnsen}, or a high-frequency decomposition of bath modes \cite{Mario Gaspar Quarenta}. According to the relativistic Dirac equation, spin inertia arises from the $1/c^4$ order spin-orbit coupling \cite{Ritwik Mondal PRB96,Ritwik Mondal JPCM30}. Through Schwinger-Keldysh field theory, the optically induced magnetic inertia term is predicted \cite{Felipe Reyes-Osorio}.

\setlength{\parskip}{0pt}

Apart from exploring the origin of magnetic inertia, substantial efforts have been invested in the inertial magnetic dynamics. Significant magnetic nutation has been evidenced via exact analytic solutions of inertial Landau-Lifshitz equation \cite{S. V. Titov PRB103}, numerical simulations of the inertial LLG equation for several nanostructured systems \cite{D. Bottcher}, and time-dependent equilibrium correlation-function analyses \cite{Sergei V. Titov PRB107}. Moreover, existing studies demonstrate that magnetic inertia gives rise to emergent nutational resonances alongside conventional ferromagnetic \cite{E. Olive APL,E. Olive JAP,Mikhail Cherkasskii PRB102,Ritwik Mondal JPCM33,Sergei V. Titov JAP,Mikhail Cherkasskii PRB106,Subhadip Ghosh PRB,Jonas Wiemeler} and antiferromagnetic \cite{Ritwik Mondal PRB103,Ritwik Mondal PRB10410,David Angster} ones. Besides these intrinsic inertial resonances, the interplay between nutational resonance and spin current has also been explored, covering nutation-injected spin pumping \cite{Ritwik Mondal PRB10421} and forced nutational resonance driven by spin torques \cite{Peng-Bin He PRB112}. Beyond the foregoing linear inertial magnetic dynamics, magnetic inertia also affects the nonlinear magnetic dynamics. Specifically, it alters the self-oscillation of magnetization in both antiferromagnets (AFMs) \cite{Peng-Bin He PRB108,Peng-Bin He PRB11006} and ferromagnets (FMs) \cite{Rodolfo Rodriguez}, and enables ultrafast magnetization switching \cite{Rahnuma Rahman,Kumar Neeraj PRB,I. Makhfudz PRB,Lucas Winter}. In previously mentioned works, however, the polarization and chirality of resonant modes have received only limited attention. They are only briefly discussed in Ref. [\onlinecite{Subhadip Ghosh PRB}] and systematically explored for forced inertial resonance \cite{Peng-Bin He PRB112}. In addition, the chirality is used to distinguish the spin pumping currents generated by nutational and precessional resonances \cite{Ritwik Mondal PRB10421}.

\setlength{\parskip}{0pt}

The aforementioned theoretical studies have focused primarily on the temporal dynamics of magnetization, while works on inertial spin waves have additionally accounted for its spatial variation \cite{I. Makhfudz APL,Mikhail Cherkasskii PRB103,Alexey M. Lomonosov,Sergei V. Titov PRB105,Ritwik Mondal PRB106,Mikhail Cherkasskii PRB109,Peng-Bin He PRB110,Massimiliano d'Aquino,H. Y. Yuan,Subhadip Ghosh JPCM}. To the best of our knowledge, only a few studies have addressed spin-wave spectra across the entire Brillouin zone \cite{Subhadip Ghosh JPCM,Mikhail Cherkasskii PRB109}, whereas most works remain restricted to the long-wavelength regime. With regards to the chirality, Ref. [\onlinecite{Alexey M. Lomonosov}] points out that the magnetization precesses in opposite directions for the nutational and precessional SWs in ferromagnets. Furthermore, nearly all of these spin-wave analyses assume uniform ferromagnetic or antiferromagnetic ground states, while spin waves on top of nonuniform configurations appear to have been addressed only in very few works, for instance, the inertial SWs on a spin spiral \cite{Mikhail Cherkasskii PRB109}.

While inertial SWs (and especially their dispersions) in uniform FMs and AFMs are well investigated, those in more complicated structures hosting spatially textured magnetization, such as canted and spiral configurations, are less explored. Moreover, a systematic characterization of the chirality and polarization of inertial SWs is still lacking, and the connection between spin-wave chirality and the sign of the eigenfrequency remains ambiguous.
Therefore, we develop a more straightforward formalism for calculating inertial SWs, applicable to both uniform and nonuniform magnetic configurations. Furthermore, we propose a general definition of the chirality and polarization of SWs. The formalism is validated through a case study of uniaxial AFMs with staggered and homogeneous Dzyaloshinskii-Moriya interactions (DMIs).

This paper is structured as follows. Section \ref{method} outlines the theoretical framework adopted throughout our analysis. Sec. \ref{equilibrium configurations} details the equilibrium configurations of uniaxial AFMs with staggered and homogeneous DMIs. In Sec. \ref{Inertial SWs in AFM with staggered DMI}, we investigate inertial SWs in AFMs with staggered DMI, covering excitations built upon uniform AFM and canted equilibrium configurations. The inertial SWs for homogeneous-DMI AFMs are given in Sec. \ref{Inertial SWs in AFM with homogeneous DMI}, where we address SWs originating from uniform AFM and spiral ground states. Finally, some discussions and conclusions are presented in Secs. \ref{Discussions} and \ref{Conclusions}, respectively. Some complicated intermediate formulas and derivations are provided in the Appendix.

\vspace{-1em}

\section{Methodology} \label{method}

The discrete inertial Landau-Lifshitz (ILL) equation reads
\begin{equation}
\frac{\partial \mathbf{m}_l}{\partial t} = - \mathbf{m}_l \times \mathbf{h}^{eff}_l + \eta \mathbf{m}_l \times \frac{\partial^2 \mathbf{m}_l}{\partial t^2}, \label{reduced ILL equations}
\end{equation}
where $\eta$ is the inertial relaxation time, $\mathbf{m}_l$ is the unit vector of the magnetization $\mathbf{M}_l$ at $l$-th site. The effective field $\mathbf{h}^{eff}_l$ can be calculated by $\mathbf{h}^{eff}_l = - \partial E/\partial \mathbf{m}_l$, with the reduced magnetic energy $E$ having the dimension of frequency; see for example Eq. (\ref{reduced magnetic energy}).

 Most static magnetic configurations are spatially nonuniform. To analyze SWs living on top of them, it is convenient to take the local sets of coordinates attached to the spatially varying equilibrium magnetization. The unit basis vectors of the local frame are defined as
\begin{equation}
\! \left( \! \begin{array}{c} \mathbf{e}_r^l \\[1.5mm] \mathbf{e}_\theta^l \\[1.5mm] \mathbf{e}_\phi^l \end{array} \! \right) \!
= \! \left( \! \begin{array}{ccc} \sin \theta^0_l \cos \phi^0_l & \sin \theta^0_l \sin \phi^0_l & \cos \theta^0_l \\[1.5mm]
\cos \theta^0_l \cos \phi^0_l & \cos \theta^0_l \sin \phi^0_l & - \sin \theta^0_l \\[1.5mm] - \sin \phi^0_l & \cos \phi^0_l & 0 \end{array} \! \right) \!
\left( \! \begin{array}{c} \mathbf{e}_x \\[1.5mm] \mathbf{e}_y \\[1.5mm] \mathbf{e}_z \end{array} \! \right) \!, \label{local frame}
\end{equation}
with $\theta^0_l$  being the polar angle between $\mathbf{m}^0_l$ and the positive $z$-axis, and $\phi^0_l$ being the azimuthal angle between the projection of $\mathbf{m}^0_l$ on the $x y$-plane and the $x$-axis.
Here, $\mathbf{m}^0_l$ denotes the equilibrium direction of the magnetization at the $l$-th site and satisfies the torque-free condition $\mathbf{m}_l \times \mathbf{h}^{eff}_l \vert_{\mathbf{m}_l = \mathbf{m}_l^0} = 0$. By this definition and Eq. (\ref{local frame}), $\mathbf{e}^l_r = \mathbf{m}^0_l$.

The spin-wave ansatz is assumed as
\begin{equation}
\mathbf{m}_l = \mathbf{m}_l^0 + \delta \mathbf{m}_l, \label{spin-wave ansatz}
\end{equation}
where $\delta \mathbf{m}_l$ is the dynamic components, which is small in the linear regime. In the local coordinate frame defined in Eq. (\ref{local frame}), the spin-wave fluctuation can be expanded up to second order in the small quantities as
\begin{equation}
\delta \mathbf{m}_l \approx m_{\theta, l} \mathbf{e}_\theta^l + m_{\phi, l} \mathbf{e}_\phi^l - \frac{1}{2} \left( m^2_{\theta, l} + m^2_{\phi, l} \right) \mathbf{e}^l_r, \label{spin-wave fluctuation}
\end{equation}
with $m_{\theta (\phi), l}$ being the component of spin-wave fluctuation in $\mathbf{e}_{\theta  (\phi)}^l$ direction. According to Eq. (\ref{spin-wave fluctuation}), the reduced energy is a function of $m_{\theta, l}$ and $m_{\phi, l}$. The effective field in the local frame can be calculated by
\begin{equation}
\mathbf{h}^{eff}_l = - \frac{\partial E}{\partial m_{\theta, l}} \mathbf{e}_\theta^l - \frac{\partial E}{\partial m_{\phi, l}} \mathbf{e}_\phi^l. \label{effective field}
\end{equation}
Remaining the zero-order and linear terms of $m_{\theta, l}$ and $m_{\phi, l}$, the effective field are expanded as
\begin{equation}
\mathbf{h}^{eff}_l = \mathbf{h}^0_l + \delta \mathbf{h}_l, \label{linearized effective field}
\end{equation}
where $\mathbf{h}^0_l$ is the effective field in equilibrium, and $\delta \mathbf{h}_l$ is the linear fluctuation of the effective field. The expressions of $\mathbf{h}^0_l$ and $\delta \mathbf{h}_l$ are presented in Appendix. \ref{expansion of effective field}.

Substituting the spin-wave ansatz described by Eq. (\ref{spin-wave ansatz}) and (\ref{spin-wave fluctuation}) and the effective field expansion Eq. (\ref{linearized effective field}) into the ILL equation (\ref{reduced ILL equations}), and utilizing the equilibrium equation $\mathbf{m}_l^0 \times \mathbf{h}^0_l = 0$, one can get a set of differential equations for $(m_{\theta, l}, m_{\phi, l}$) after linearization,
\begin{eqnarray}
\! && \! \frac{\partial}{\partial t} \! \left( \! \begin{array}{c} m_{\theta, l} \\ m_{\phi, l} \end{array} \! \right) \! + \eta \Omega \frac{\partial^2}{\partial t^2} \! \left( \! \begin{array}{c} m_{\theta, l} \\ m_{\phi, l} \end{array} \! \right) \! = - \Omega \times \notag \\[0.5em]
\! && \! \left[ \! \mathcal{H}_l \! \left( \! \begin{array}{c} m_{\theta, l} \\ m_{\phi, l} \end{array} \! \right) \! + \! \mathcal{H}_{l + 1} \! \left( \! \begin{array}{c} m_{\theta, l + 1} \\ m_{\phi, l + 1} \end{array} \! \right) \! + \! \mathcal{H}_{l - 1} \! \left( \! \begin{array}{c} m_{\theta, l - 1} \\ m_{\phi, l - 1} \end{array} \! \right) \! \right] \!, \label{linearized equation}
\end{eqnarray}
where $\Omega$ is the symplectic unit matrix,
\begin{equation}
\Omega = \left( \begin{array}{cc} 0 & 1 \\ - 1 & 0 \end{array} \right). \label{symplectic unit matrix}
\end{equation}
The three $\mathcal{H}$ matrixes read
\begin{equation}
\mathcal{H}_{l^\prime} = \left( \begin{array}{cc}\frac{\partial^2 E}{\partial m_{\theta, l} \partial m_{\theta, l^\prime}} & \frac{\partial^2 E}{\partial m_{\theta, l} \partial m_{\phi, l^\prime}} \\[1em] \frac{\partial^2 E}{\partial m_{\phi, l} \partial m_{\theta, l^\prime}} & \frac{\partial^2 E}{\partial m_{\phi, l} \partial m_{\phi, l^\prime}} \end{array} \right)_0, \label{Hessian matrix}
\end{equation}
where, the subscript `$0$' indicates that the derivatives are evaluated at the equilibrium state for $l^\prime = l, l + 1, l - 1$. Here, the magnetic energy is restricted to nearest-neighbor interactions, giving rise to coupling between site $l$ and its adjacent sites $l \pm 1$.

Eq. (\ref{linearized equation}) constitutes a coupled system of linear second-order differential equations, which are generally site-dependent. Introducing the auxiliary phase-space variables $\partial m_{\theta, l}/\partial t$ and $\partial m_{\phi, l}/\partial t$, the original second-order system can be recast as a first-order one with doubled dimensionality. This reformulation significantly simplifies the evaluation of spin-wave eigenfrequencies and their corresponding eigenvectors. The resulting first-order differential equations can be written as
\begin{equation}
\frac{\partial \mathsf{v}_l}{\partial t} = - i \mathcal{M}_l \mathsf{v}_l - i \mathcal{M}_{l + 1} \mathsf{v}_{l + 1} - i \mathcal{M}_{l - 1} \mathsf{v}_{l - 1}. \label{first-order differential equation}
\end{equation}
After introducing two new dynamic variables, the vectors $\mathsf{v}_{l}$ are defined as
\begin{equation}
\mathsf{v}_{l} = \left( m_{\theta, l}, m_{\phi, l}, i \frac{\partial m_{\theta, l}}{\partial t}, i \frac{\partial m_{\phi, l}}{\partial t} \right)^T, \label{dynamic vector}
\end{equation}
with $T$ denoting the transpose of matrix. Vectors $\mathsf{v}_{l \pm 1}$ are defined in the identical manner. The top half part of $\mathsf{v}_l$ gives the linear SW.  The parameter matrixes are also expanded as
\begin{equation}
\mathcal{M}_l = \left( \begin{array}{cc} O_2 & I_2 \\[1em] \frac{1}{\eta} \mathcal{H}_l & \frac{i}{\eta} \Omega \end{array} \right), \label{parameter matrix at site l}
\end{equation}
and
\begin{equation}
\mathcal{M}_{l \pm 1} = \left( \begin{array}{cc} O_2 & O_2 \\[1em] \frac{1}{\eta} \mathcal{H}_{l \pm 1} & O_2 \end{array} \right). \label{parameter matrix at site l +- 1}
\end{equation}
with $O_2$ and $I_2$ being the $2 \times 2$ zero matrix and identity matrix, respectively.

Assuming the plane-wave solution, we have
\begin{equation}
\left( \! \begin{array}{c} m_{\theta, l} \\[0.5em] m_{\phi, l} \end{array} \! \right) = \left( \! \begin{array}{c} \mathcal{A}_{\theta, l} \\[0.5em] \mathcal{A}_{\phi, l} \end{array} \! \right) e^{- i \omega t + i k l a}, \label{plane-wave solution}
\end{equation}
with $k$ being the wave vector and $a$ the lattice constant. $\mathcal{A}_{\theta (\phi), l}$ is the complex amplitude of $m_{\theta (\phi), l}$. $m_{\theta (\phi), l \pm 1}$ follows the same form. Inserting Eq. (\ref{plane-wave solution}) into Eqs. (\ref{dynamic vector}) and (\ref{first-order differential equation}) yields the spin-wave eigen-equation,
\begin{equation}
\widetilde{\mathcal{M}}_l \left( \! \begin{array}{c} \mathcal{A}_{\theta, l} \\[0.5em] \mathcal{A}_{\phi, l} \\[0.5em] \omega \mathcal{A}_{\theta, l} \\[0.5em] \omega \mathcal{A}_{\phi, l} \end{array} \! \right) = \omega \! \left( \! \begin{array}{c} \mathcal{A}_{\theta, l} \\[0.5em] \mathcal{A}_{\phi, l} \\[0.5em] \omega \mathcal{A}_{\theta, l} \\[0.5em] \omega \mathcal{A}_{\phi, l} \end{array} \! \right), \label{spin-wave eigen-equation}
\end{equation}
where $\widetilde{\mathcal{M}}_l = \mathcal{M}_l + e^{i k a} \mathcal{M}_{l + 1} + e^{- i k a} \mathcal{M}_{l - 1}$. Eq. (\ref{spin-wave eigen-equation}) and its matrixes Eqs. (\ref{parameter matrix at site l}), (\ref{parameter matrix at site l +- 1}) and (\ref{Hessian matrix}) constitute the core of our method. The solution of this system provides the inertial spin-wave modes associated for various uniform and nonuniform magnetic configurations.

After getting the explicit expressions of $\mathcal{A}_{\theta, l}$ and $\mathcal{A}_{\phi, l}$, the precessional trajectories of magnetization can be reconstructed from Eq. (\ref{plane-wave solution}) by taking the real part of $m_{\theta,l}(t)$ and $m_{\phi,l}(t)$. The plane-wave solution of SW reads
\begin{equation}
m_{\theta (\phi), l} = A_{\theta (\phi), l} \cos \left( - \omega t + k l a + \delta_{\theta (\phi), l} \right). \label{components of spin-wave fluctuation}
\end{equation}
In Eq. (\ref{components of spin-wave fluctuation}), $A_{\theta (\phi), l}$ denotes the real amplitude derived from the modulus of $\mathcal{A}_{\theta (\phi), l}$, while $\delta_{\theta (\phi), l}$ represents the initial phase corresponding to the argument of $\mathcal{A}_{\theta (\phi), l}$.

Here, we define the chirality as the rotational sense of $\mathbf{m}_l$ around its equilibrium magnetization, i.e. $\mathbf{e}_r^l = \mathbf{m}^0_l$. It is convenient to represent both the chirality and polarization of a SW by using a single parameter, known as the ellipticity angle \cite{D. H. Goldstein}, which is defined as
\begin{equation}
\sin 2 \chi_l = \text{sgn}(\omega) \frac{2 A_{\theta, l} A_{\phi, l}}{A_{\theta, l}^2 + A_{\phi, l}^2} \sin \delta, \label{ellipticity angle}
\end{equation}
where $\delta = \delta_{\phi, l} - \delta_{\theta, l}$ denotes the phase difference and $\operatorname{sgn} (\omega) = \omega/\vert \omega \vert$. The ellipticity angle is restricted to the interval $\chi_l \in [ - \pi/4, \pi/4]$.
For $\vert \chi \vert = 0$, the spin wave is linearly polarized, with the polarization direction lying anywhere in the local transverse plane, including along the $\mathbf{e}_{\theta}$ or $\mathbf{e}_{\phi}$ axes as special cases. For $\vert \chi_l \vert = \pi/4$, the SW is circularly polarized. For $0 < \vert \chi \vert < \pi/4$, the SW is elliptically polarized. As $\vert \chi \vert$ increases, the polarization approaches more circular. For positive (negative) $\chi$, the chirality is righthanded (lefthanded). It should be emphasized that only the sign of $\omega$ can not determine the chirality of a SW, which also depends on the phase difference $\delta$ between two dynamic components of magnetization. Of course, flipping the sign of $\omega$ results in the reversal of chirality if $\sin \delta$ preserves its sign.

To demonstrate the robustness of the proposed formalism, we apply it to calculate the dispersion relation, polarization, chirality, and phase of inertial SWs. The studied systems are uniaxial antiferromagnets with both homogeneous and staggered DMI, hosting either uniform or textured magnetic configurations. We first define their equilibrium configurations.

\vspace{-1em}

\section{equilibrium configurations} \label{equilibrium configurations}

\begin{figure*}[!t]
\includegraphics[clip, trim = 0cm 0cm 0cm 0cm, width=0.8\textwidth]{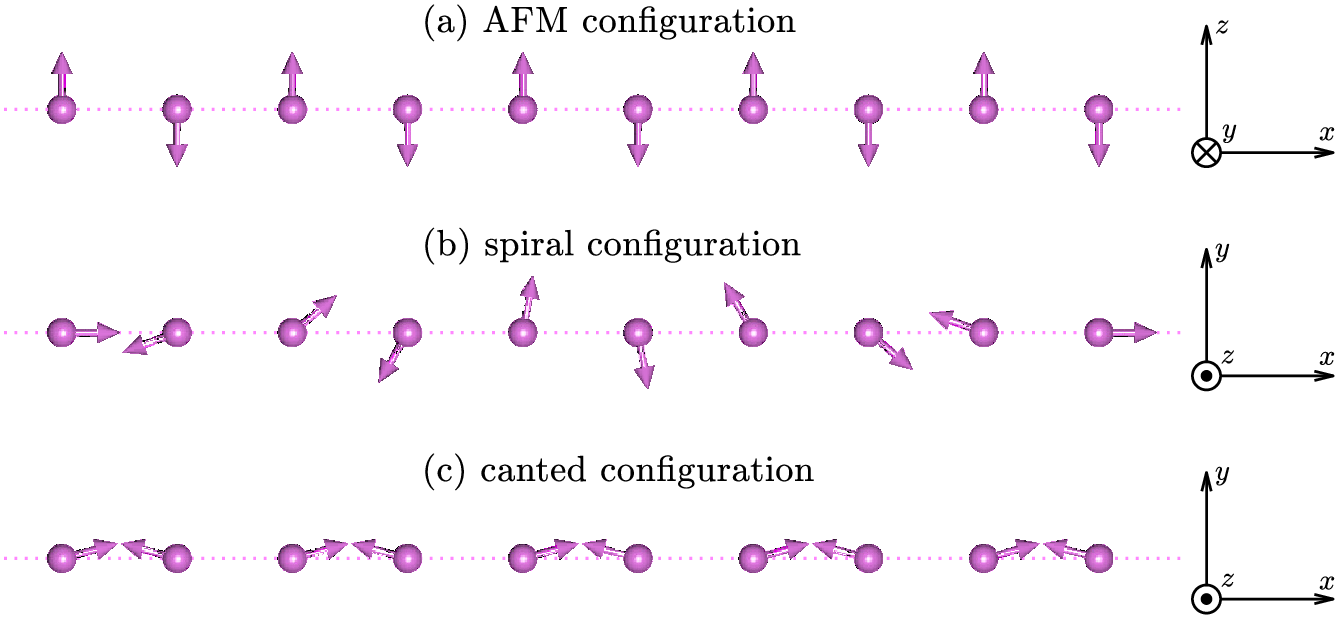}
\caption{(color online). The static magnetic configurations of AFMs with homogeneous and staggered DMIs.}
\label{magnetic configurations}
\end{figure*}

The magnetic energy under consideration accounts for the contributions from the exchange interaction, the uniaxial magnetic anisotropy, and the DMI. Expressed in the frequency unit, the reduced magnetic energy reads
\begin{eqnarray}
E &=& \sum_l \left[ \omega_E \mathbf{m}_l \cdot \mathbf{m}_{l + 1} - \omega_K \left( \mathbf{m}_l \cdot \mathbf{e}_z \right)^2 \right. \notag \\ && + \lambda \omega_D \mathbf{e}_z \cdot \left( \mathbf{m}_l \times \mathbf{m}_{l + 1} \right) \Big], \label{reduced magnetic energy}
\end{eqnarray}
where $\omega_E = \gamma \mathsf{J}/(\mu_0 M)$, $\omega_K = \gamma \mathsf{K}/(\mu_0 M)$, and $\omega_D = \gamma \mathsf{D}/\mu_0 M)$, with $\mathsf{J} > 0$ being the exchange energy density between the nearest local magnetic moments, $\mathsf{K}$ the easy-axis anisotropy constant, $\mathsf{D}$ the DM constant, and $M$ the magnitude of $\mathbf{M}_l$. The choice $\lambda = 1$ corresponds to homogeneous DMI, whereas $\lambda = (- 1)^l$ describes staggered DMI \cite{I. Dzyaloshinskii,Toru Moriya,J. H. H. Perk,Carmine Autieri,Xiyin Ye}. Here, it is assumed that the easy axis and the DM vector are along the $z$-axis. Comparing with the homogeneous DMI, only an additional factor $(- 1)^l$ appears in the staggered DMI term. This modification implies that the staggered DM vectors for neighboring magnetic moments align in opposite directions.

Minimizing the magnetic energy [eq. (\ref{reduced magnetic energy})] determines the static magnetic configuration. In equilibrium, the system adopts the AFM configuration shown in Fig. \ref{magnetic configurations}(a) when $\vert \omega_D \vert < \omega^c_D$. The critical value of $\omega_D$ reads
\begin{equation}
\omega^c_D = \sqrt{\omega_K (2 \omega_E + \omega_K)}. \label{critical DMI}
\end{equation}
When $\vert \omega_D \vert > \omega^c_D$, the system twists into a spiral state for homogeneous DMI \cite{J. H. H. Perk,Xiyin Ye}, as shown in Fig. \ref{magnetic configurations}(b), whereas for staggered DMI \cite{J. H. H. Perk,Carmine Autieri,Xiyin Ye} it enters the canted configuration shown in Fig. \ref{magnetic configurations}(c). In the absence of anisotropy, one obtains $\omega^c_D = 0$, such that homogeneous DMI immediately drives the system into a spiral state, whereas staggered DMI gives rise to a canted configuration.

The spiral state is described by $m^z_l = 0$, $m^x_l = \cos (l \phi_s)$, $m^y_l = \sin (l \phi_s)$, with spin rotation angle per lattice step
\begin{equation}
\phi_s = \tan^{- 1} \left( - \omega_E, - \omega_D \right), \label{spiral angle}
\end{equation}
where $\operatorname{\tan}^{- 1}(x, y)$ denotes the two-argument arctangent that returns the angle of the vector $(x,y)$, correctly resolving the quadrant \cite{arctan}. Evidently, for $\omega_E > 0$, $\phi_s$ is close to a straight angle owing to $\omega_D \ll \omega_E$.

The canted state is double degenerate, described by $m^z_l = 0$, $m^x_{2 j + 1} = - m^x_{2 j} = \pm \cos (\phi_c)$, $m^y_{2 j + 1} = m^y_{2 j} = \pm \sin (\phi_c)$, with the canted angle
\begin{equation}
\phi_c = \frac{1}{2} \tan^{- 1} \left( \omega_E, \omega_D \right). \label{canted angle}
\end{equation}

Based on the foregoing AFM, spiral and canted configurations, we proceed to compute the SWs below.

\vspace{-1em}

\section{Inertial spin waves in AFM with staggered DMI} \label{Inertial SWs in AFM with staggered DMI}

\subsection{Spin waves on top of AFM configuration} \label{secIVA}

When $\omega_D < \omega_D^c$, the AFM configuration serves as the equilibrium state. We derive the spin-wave solutions using the approach introduced in Sec. \ref{method}, with the detailed derivations provided in Appendix \ref{Spin-wave solution in staggered-DMI antiferromagnetic configuration}. For two sublattices, Eq. (\ref{components of spin-wave fluctuation}) is recast as
\begin{align}
m_{\theta (\phi), 2j} \! &= \! A_{\theta (\phi), 2j} \cos \bigl[ - \omega t \! + \! 2 j k a \! + \! \delta_{\theta (\phi), 2j} \bigr], \label{components of spin-wave fluctuation for 2j sublattice} \\
m_{\theta (\phi), 2j+1} \! &= \! A_{\theta (\phi), 2j+1} \cos \bigl[  - \omega t \! + \! (2j \! + \! 1)k a \! + \! \delta_{\theta (\phi), 2j+1} \bigr]. \label{components of spin-wave fluctuation for 2j+1 sublattice}
\end{align}

\vspace{-1em}

\begin{figure}[b]
\includegraphics[clip, trim = 0cm 2cm 0cm 2cm, width=0.5\textwidth]{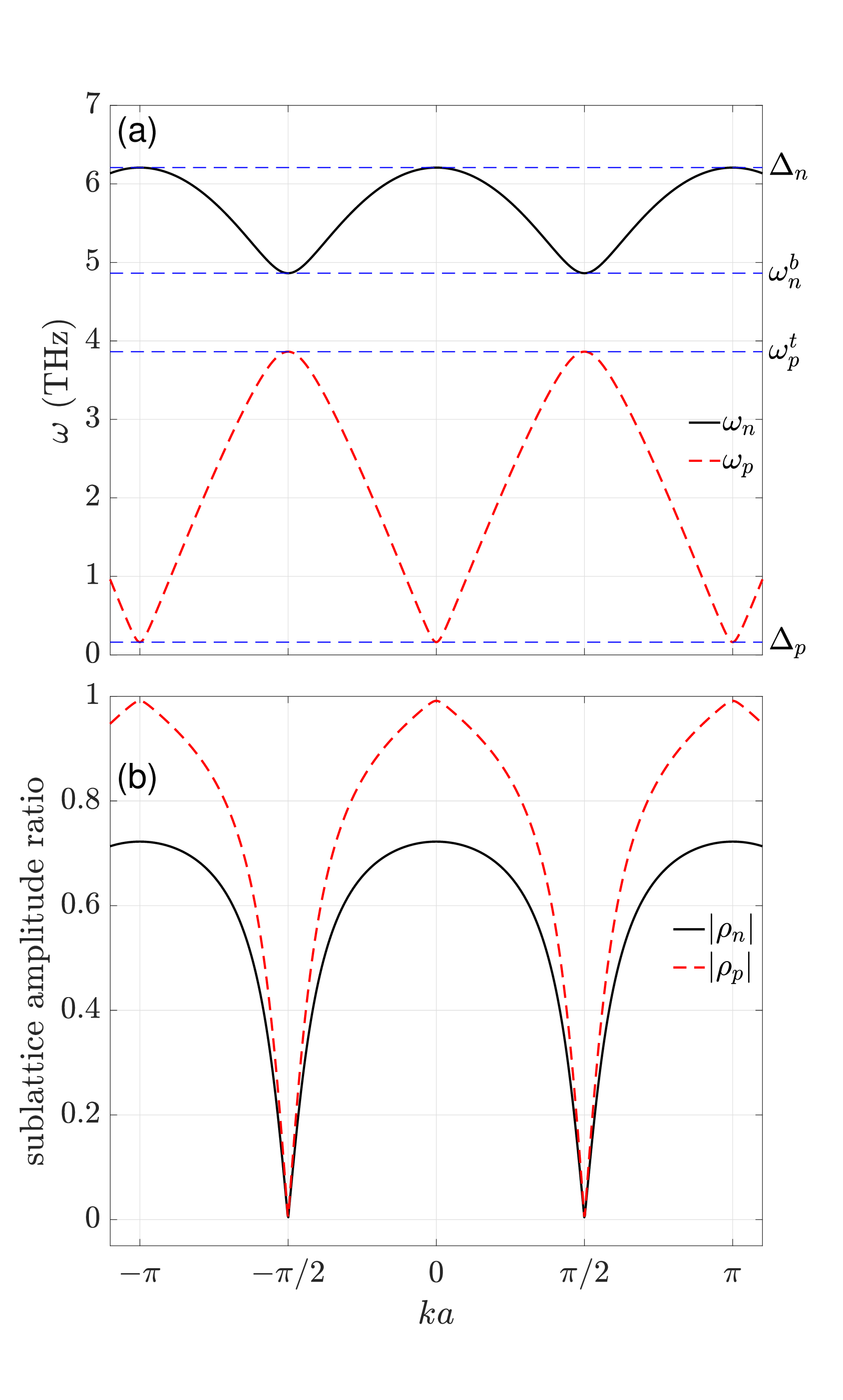}
\caption{(color online). Dispersion (a) and sublattice amplitude ratio (b) of the SWs on top of an AFM configuration for the case of staggered DMI. Here, $\omega_E = 9.25$ THz and $\omega_K = 0.14$ THz, which are derived from the magnetic parameters of MnF$_2$ \cite{Priyanka Vaidya}. The inertial relaxation time $\eta$ is set to $1$ ps. $\omega_D$ is taken as $\frac{19}{20} \omega_D^c$, where $\omega_D^c \approx 1.62$ THz is evaluated from Eq. (\ref{critical DMI}).}
\label{dispersion and amplitude ratio of staggered DMI-AFM configuration}
\end{figure}

In this case, DMI fails to lift the degeneracy. The spin-wave spectrum consists of four branches: two positive-frequency branches and their two negative-frequency counterparts. Each branch is doubly degenerate. Hence, numerous solutions can be constructed through distinct linear combinations of degenerate mode pairs. In particular, circularly polarized SWs have been extensively investigated \cite{Sergio M. Rezende}. By linearly superposing the degenerate modes, we obtain such circularly polarized SWs. Their amplitudes and initial phases, as well as the ellipticity angles [calculated from Eq. (\ref{ellipticity angle})] are listed in Tab. \ref{spin-wave parameters of staggered DMI-AFM configuration} for every branches of SWs. To better clarify these characteristics, we illustrate the precessional trajectories of the magnetization tip for all modes of the two sublattices in Tab. \ref{schematic modes of staggered DMI-AFM configuration}. This schematic clearly presents their chirality, amplitude ratios and phase differences. Additionally, we plot the dispersions and the dependence of sublattice amplitude ratios on the wave number in Fig. \ref{dispersion and amplitude ratio of staggered DMI-AFM configuration}.

\begin{table*}[!t]
\caption{Amplitudes, phases, and ellipticity angles of the SWs on top of an AFM configuration for staggered DMI. Here, $\text{H}(\lambda)$ is the unit step function and $\lambda = \sgn (\cos k a)$. To better compare the amplitude ratio between two sublattices, all the amplitudes of mode II of nutation (precession) are multiplied by $\vert \rho_n \vert$($\vert \rho_p \vert$). Here, $\varphi_0 = \arctan(\omega_E, \omega_D)$.}
\begin{tblr}{colsep=4.5pt,
  colspec = {*{13}{c}},
  rows = {valign = m}
}
\hline \hline
\SetCell[r=2,c=1]{c} {} & & \SetCell[c=2]{c} {$\omega_n$} & & & \SetCell[c=2]{c} {$- \omega_n$} & & & \SetCell[c=2]{c} {$\omega_p$} & & & \SetCell[c=2]{c} {$- \omega_p$} \\
\cline{3-4} \cline{6-7} \cline{9-10} \cline{12-13}
& & mode I & mode II & & mode I & mode II & & mode I & mode II & & mode I & mode II \\
\hline
$A_{\theta, 2 j}$ & & $\frac{\left \vert \rho_n \right \vert}{\omega_n}$ & $\frac{1}{\omega_n}$ & & $\frac{\left \vert \rho_n \right \vert}{\omega_n}$ & $\frac{1}{\omega_n}$ & & $\frac{\left \vert \rho_p \right \vert}{\omega_p}$ & $\frac{1}{\omega_p}$ & & $\frac{\left \vert \rho_p \right \vert}{\omega_p}$ & $\frac{1}{\omega_p}$ \\
$A_{\phi, 2 j}$ & & $\frac{\left \vert \rho_n \right \vert}{\omega_n}$ & $\frac{1}{\omega_n}$ & & $\frac{\left \vert \rho_n \right \vert}{\omega_n}$ & $\frac{1}{\omega_n}$ & & $\frac{\left \vert \rho_p \right \vert}{\omega_p}$ & $\frac{1}{\omega_p}$ & & $\frac{\left \vert \rho_p \right \vert}{\omega_p}$ & $\frac{1}{\omega_p}$ \\
$\delta_{\theta,2 j}$ & & {\footnotesize $\pi \text{H} (- \lambda) \! - \! \varphi_0$} & {\footnotesize $\pi \text{H} (\lambda) \! + \! \varphi_0$} & & {\footnotesize $\pi \text{H} (- \lambda) \! + \! \varphi_0$} & {\footnotesize $\pi \text{H} (\lambda) \! - \! \varphi_0$} & & {\footnotesize $\pi \text{H} (- \lambda) \! + \! \varphi_0$} & {\footnotesize $\pi \text{H} (\lambda) \! - \! \varphi_0$} & & {\footnotesize $\pi \text{H} (- \lambda) \! - \! \varphi_0$} & {\footnotesize $\pi \text{H} (\lambda) \! + \! \varphi_0$} \\
$\delta_{\phi,2 j}$ & & {\footnotesize $\lambda \frac{\pi}{2} \! - \! \varphi_0$} & {\footnotesize $\lambda \frac{\pi}{2} \! + \! \varphi_0$} & & {\footnotesize $- \lambda \frac{\pi}{2} \! + \! \varphi_0$} & {\footnotesize $- \lambda \frac{\pi}{2} \! - \! \varphi_0$} & & {\footnotesize $- \lambda \frac{\pi}{2} \! + \! \varphi_0$} & {\footnotesize $- \lambda \frac{\pi}{2} \! - \! \varphi_0$} & & {\footnotesize $\lambda \frac{\pi}{2} \! - \! \varphi_0$} & {\footnotesize $\lambda \frac{\pi}{2} \! + \! \varphi_0$} \\
\hline
$A_{\theta, 2 j + 1}$ & & $\frac{1}{\omega_n}$ & $\frac{\left \vert \rho_n \right \vert}{\omega_n}$ & & $\frac{1}{\omega_n}$ & $\frac{\left \vert \rho_n \right \vert}{\omega_n}$ & & $\frac{1}{\omega_p}$ & $\frac{\left \vert \rho_p \right \vert}{\omega_p}$ & & $\frac{1}{\omega_p}$ & $\frac{\left \vert \rho_p \right \vert}{\omega_p}$ \\
$A_{\phi, 2 j + 1}$ & & $\frac{1}{\omega_n}$ & $\frac{\left \vert \rho_n \right \vert}{\omega_n}$ & & $\frac{1}{\omega_n}$ & $\frac{\left \vert \rho_n \right \vert}{\omega_n}$ & & $\frac{1}{\omega_p}$ & $\frac{\left \vert A_p \right \vert}{\omega_p}$ & & $\frac{1}{\omega_p}$ & $\frac{\left \vert \rho_p \right \vert}{\omega_p}$ \\
$\delta_{\theta,2 j + 1}$ & & 0 & $\pi$ & & 0 & $\pi$ & & $\pi$ & $0$ & & $\pi$ & $0$ \\
$\delta_{\phi,2 j + 1}$ & & $- \frac{\pi}{2}$ & $- \frac{\pi}{2}$ & & $\frac{\pi}{2}$ & $\frac{\pi}{2}$ & & $- \frac{\pi}{2}$ & $- \frac{\pi}{2}$ & & $\frac{\pi}{2}$ & $\frac{\pi}{2}$ \\
\hline
$\chi_{2 j}$ & & $\frac{\pi}{4}$ & $- \frac{\pi}{4}$ & & $\frac{\pi}{4}$ & $- \frac{\pi}{4}$ & & $- \frac{\pi}{4}$ & $\frac{\pi}{4}$ & & $- \frac{\pi}{4}$ & $\frac{\pi}{4}$ \\
$\chi_{2 j + 1}$ & & $- \frac{\pi}{4}$ & $\frac{\pi}{4}$ & & $- \frac{\pi}{4}$ & $\frac{\pi}{4}$ & & $\frac{\pi}{4}$ & $- \frac{\pi}{4}$ & & $\frac{\pi}{4}$ & $- \frac{\pi}{4}$ \\ \hline \hline
\end{tblr}
\label{spin-wave parameters of staggered DMI-AFM configuration}
\end{table*}

In Tab. \ref{spin-wave parameters of staggered DMI-AFM configuration}, the eigenfrequencies of nutational and precessional SWs read
\begin{equation}
\omega_{n,p} = \frac{\sqrt{1 + \eta \Gamma \pm \sqrt{\left( 1 + \eta \Gamma \right)^2 - \eta^2 \Lambda^2}}}{\sqrt{2} \eta}, \label{omega of staggered DMI-AFM configuration}
\end{equation}
where $\Gamma = 4 \sqrt{\omega_E^2 + \omega_D^{c 2}}$, and $\Lambda = 4 [ ( \omega_E^2 + \omega_D^{c 2} ) - ( \omega_E^2 + \omega_D^2 ) \cos^2 k a ]^{1/2}$. In Eq. (\ref{omega of staggered DMI-AFM configuration}), the subscripts ${n,p}$ refer, respectively, to the upper and lower signs of $\pm$ in front of the square root. In the limit that $\eta \rightarrow 0$, $\omega_n$ approaches infinity, and $\omega_p$ becomes $2 [ ( \omega_E^2 + \omega_D^{c 2} ) - ( \omega_E^2 + \omega_D^2 ) \cos^2 k a ]^{1/2}$. Notably, $\Lambda$ is purely imaginary for some values of $k$ when $\omega_D > \omega_D^c$, yielding $\Lambda^2 < 0$. Consequently, $\omega_p$ becomes complex according to Eq. (\ref{omega of staggered DMI-AFM configuration}), indicating the instability of AFM configuration.

In Tab. \ref{spin-wave parameters of staggered DMI-AFM configuration}, the parameters $\rho_{n,p}$ are given by
\begin{eqnarray}
\rho_{n,p} = \pm \frac{2 \sqrt{\omega_E^2 + \omega_D^2} \cos k a}{\eta \omega_{n,p}^2 \pm \omega_{n,p} - 2 \sqrt{\omega_E^2 + \omega_D^{c 2}}}, \label{rho of staggered DMI-AFM configuration}
\end{eqnarray}
with the upper and lower signs of $\pm$ corresponding to the subscripts $n$ and $p$ of $\rho_{n,p}$. It can be verified that $\eta \omega_n^2 + \omega_n - 2 \sqrt{\omega_E^2 + \omega_D^{c 2}} > 0$ and $- (\eta \omega_p^2 - \omega_p - 2 \sqrt{\omega_E^2 + \omega_D^{c 2}}) > 0$. So, the signs of $\rho_{n,p}$ are determined by $\cos \left( k a \right)$.

\begin{table*}[t]
\centering
\caption{Schematic diagrams of the modes. The blue solid (red dotted) circles denote the trajectories of the tips of $\mathbf{m}_{ 2 j}$ ($\mathbf{m}_{ 2 j + 1}$). The arrows on the blue solid (red dotted) circles signify the rotational senses of $\mathbf{m}_{ 2 j}$ ($\mathbf{m}_{ 2 j + 1}$) around  $- \mathbf{r}_r^{ 2 j}$ ($\mathbf{e}_r^{ 2 j + 1}$). The blue solid (red dotted) vector line represent the fluctuation $\delta \mathbf{m}_{2 j}$ ($\delta \mathbf{m}_{2 j + 1}$).}
\begin{tblr}{
  colspec = {*{12}{c}},
  rows = {valign = m}
}
\hline \hline
& \SetCell[c=2]{c} {$\omega_n$} & & & \SetCell[c=2]{c} {$- \omega_n$} & & & \SetCell[c=2]{c} {$\omega_p$} & & & \SetCell[c=2]{c} {$- \omega_p$} \\
\cline{2-3} \cline{5-6} \cline{8-9} \cline{11-12}
& mode I & mode II & & mode I & mode II & & mode I & mode II & & mode I & mode II \\
\hline
$0 < \vert k \vert a < \frac{\pi}{2}$ & \adjustbox{valign=m}{\includegraphics[clip, trim = 0cm 0cm 0cm 0cm, width=0.07\textwidth]{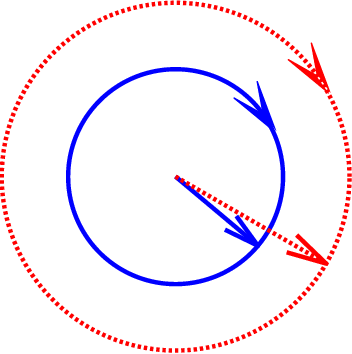}} & \adjustbox{valign=m}{\includegraphics[clip, trim = 0cm 0cm 0cm 0cm, width=0.07\textwidth]{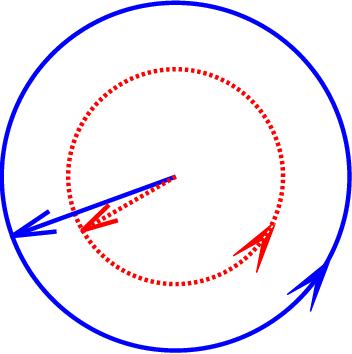}} & & \adjustbox{valign=m}{\includegraphics[clip, trim = 0cm 0cm 0cm 0cm, width=0.07\textwidth]{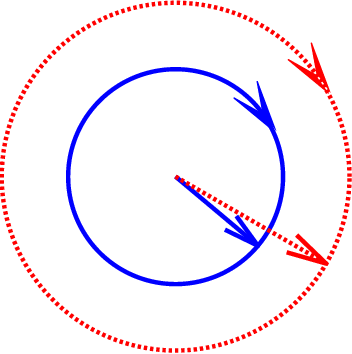}} & \adjustbox{valign=m}{\includegraphics[clip, trim = 0cm 0cm 0cm 0cm, width=0.07\textwidth]{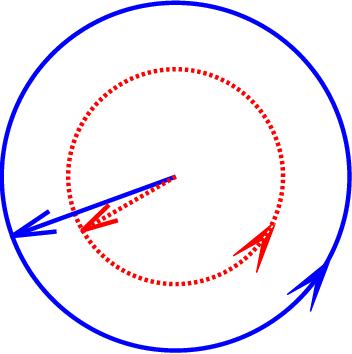}} & & \adjustbox{valign=m}{\includegraphics[clip, trim = 0cm 0cm 0cm 0cm, width=0.07\textwidth]{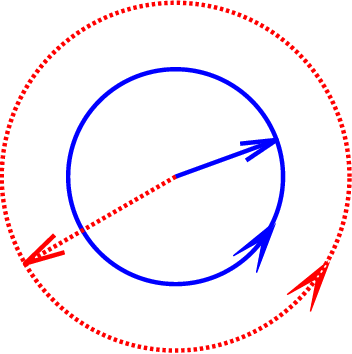}} & \adjustbox{valign=m}{\includegraphics[clip, trim = 0cm 0cm 0cm 0cm, width=0.07\textwidth]{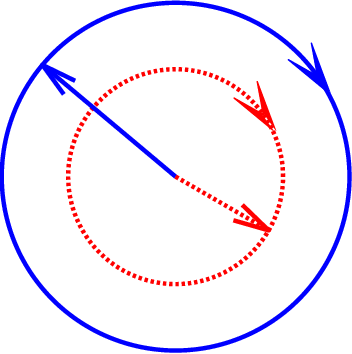}} & & \adjustbox{valign=m}{\includegraphics[clip, trim = 0cm 0cm 0cm 0cm, width=0.07\textwidth]{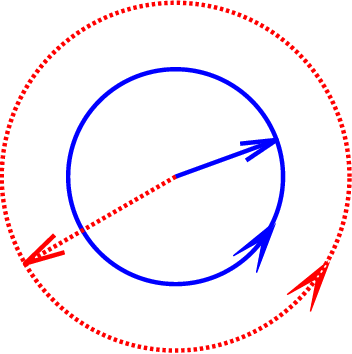}} & \adjustbox{valign=m}{\includegraphics[clip, trim = 0cm 0cm 0cm 0cm, width=0.07\textwidth]{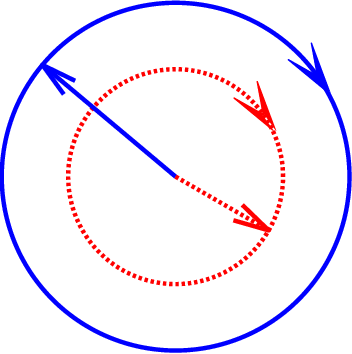}} \\
$\frac{\pi}{2} < \vert k \vert a < \pi$ & \adjustbox{valign=m}{\includegraphics[clip, trim = 0cm 0cm 0cm 0cm, width=0.07\textwidth]{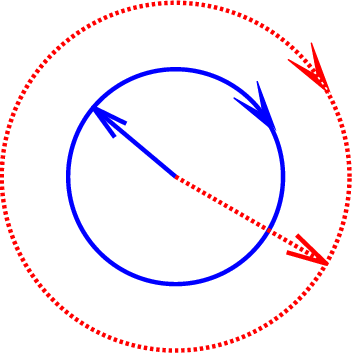}} & \adjustbox{valign=m}{\includegraphics[clip, trim = 0cm 0cm 0cm 0cm, width=0.07\textwidth]{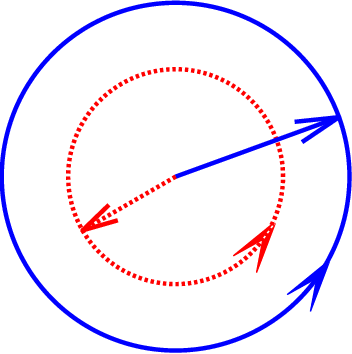}} & & \adjustbox{valign=m}{\includegraphics[clip, trim = 0cm 0cm 0cm 0cm, width=0.07\textwidth]{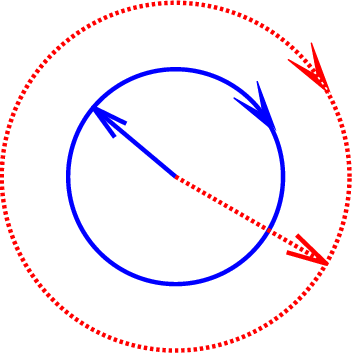}} & \adjustbox{valign=m}{\includegraphics[clip, trim = 0cm 0cm 0cm 0cm, width=0.07\textwidth]{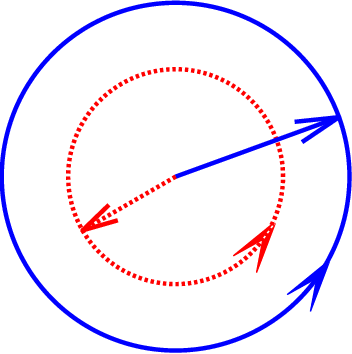}} & & \adjustbox{valign=m}{\includegraphics[clip, trim = 0cm 0cm 0cm 0cm, width=0.07\textwidth]{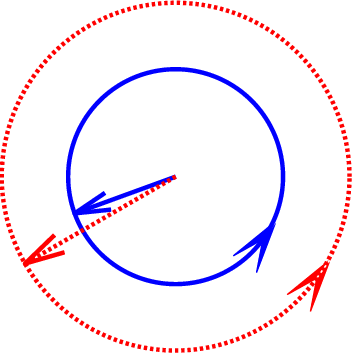}} & \adjustbox{valign=m}{\includegraphics[clip, trim = 0cm 0cm 0cm 0cm, width=0.07\textwidth]{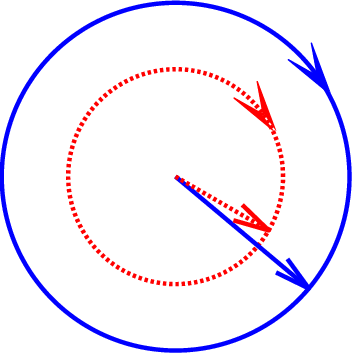}} & & \adjustbox{valign=m}{\includegraphics[clip, trim = 0cm 0cm 0cm 0cm, width=0.07\textwidth]{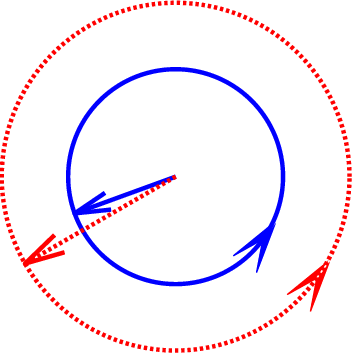}} & \adjustbox{valign=m}{\includegraphics[clip, trim = 0cm 0cm 0cm 0cm, width=0.07\textwidth]{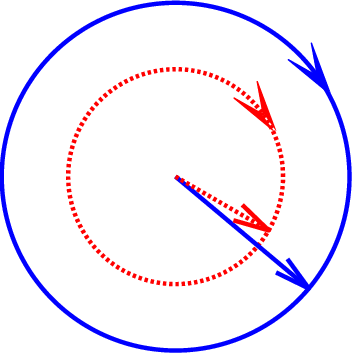}} \\
\hline
frame & \SetCell[c=2]{c}{\adjustbox{valign=m}{\includegraphics[clip, trim = 0cm 0cm 0cm 0cm, width=0.11\textwidth]{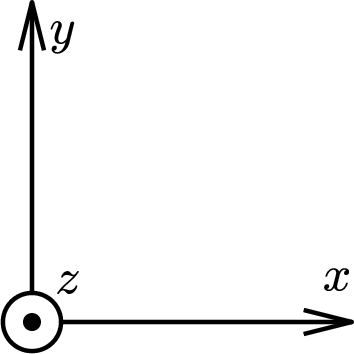}}} & & & \SetCell[c=2]{c}{\adjustbox{valign=m}{\includegraphics[clip, trim = 0cm 0cm 0cm 0cm, width=0.11\textwidth]{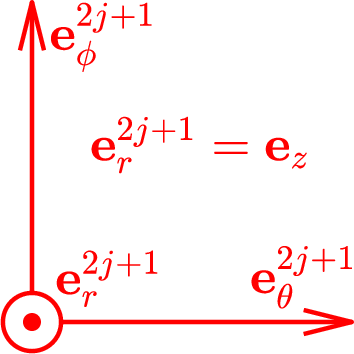}}} & & & \SetCell[c=2]{c}{\adjustbox{valign=m}{\includegraphics[clip, trim = 0cm 0cm 0cm 0cm, width=0.11\textwidth]{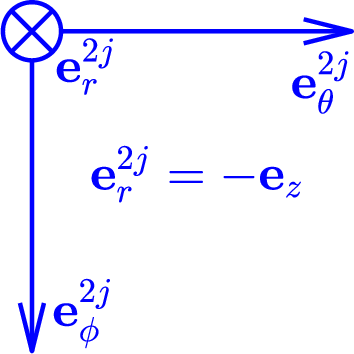}}} \\
\hline \hline
\end{tblr}
\label{schematic modes of staggered DMI-AFM configuration}
\end{table*}

\vspace{-1em}

\subsubsection{Dispersion}

Near $k = 0$, the dispersions expanded to second order in $k$ read
\begin{equation}
\omega_{n,p} = \sqrt{\Delta^2_{n,p} \pm v^2 k^2}, \label{long-wavelength dispersion of staggered DMI-AFM configuration}
\end{equation}
with the sign $\pm$ before $v^2$ corresponding to the subscripts $n$ and $p$ of $\omega_{n,p}$.
At $k = 0$, the frequency of the nutational SW reaches its maximum, as shown in Fig. \ref{dispersion and amplitude ratio of staggered DMI-AFM configuration}(a), indicating the optical-branch character of this mode. The frequency maximum of the nutational branch ($\Delta_n$) and the gap of the precessional branch ($\Delta_p$) are given by
\begin{equation}
\! \Delta_{n,p} = \frac{\sqrt{1 + \eta \Gamma \pm \! \sqrt{ \! \left( 1 + \eta \Gamma \right)^2 - 16 \eta^2 \! \left( \omega_D^{c 2} - \omega_D^2 \right)}}}{\sqrt{2} \eta}, \label{gap of staggered DMI-AFM configuration}
\end{equation}
where the subscripts $n$ and $p$ refer, respectively, to the upper and lower signs in front of the square root. The precessional spin-wave gap $\Delta_p$ is reduced by the spin inertia, as can be seen from the expansion:
\begin{equation}
\Delta_p = 2 \sqrt{\omega_D^{c 2} - \omega_D^2} \left[ 1 - \Gamma \eta + O(\eta^2) \right].
\end{equation}
When $\omega_D = \omega_D^c$, the gap vanishes; when $\omega_D > \omega_D^c$, $\Delta_p$ becomes complex, leading to the instability of these SWs.

The characteristic velocity $v$ in Eq. (\ref{long-wavelength dispersion of staggered DMI-AFM configuration}) is
\begin{equation}
v = \frac{2 \sqrt{\omega_E^2 + \omega_D^2} a}{\left[ \left( 1 + \eta \Gamma \right)^2 - 16 \eta^2 \left( \omega_D^{c 2} - \omega_D^2 \right) \right]^{\frac{1}{4}}}, \label{velocity of staggered DMI-AFM configuration}
\end{equation}
which is identical for the nutational and precessional SWs. From Eq. (\ref{velocity of staggered DMI-AFM configuration}), it can be inferred that the presence of spin inertia diminishes the velocity. Obviously, the group velocity of the nutational SW has the opposite sign to the wave number $k$, which classifies it as a backward SW.

At $k = \pm \pi/(2 a)$, the nutational spin-wave branch reaches the bottom of its dispersion band, whereas the precessional spin-wave branch reaches the top of its band. The frequencies at the bottom of nutational band and the top of precessional band read
\begin{equation}
\omega_{n(p)}^{b(t)} = \frac{\sqrt{1 + 8 \eta \left( \omega_E + \omega_K \right)} \pm 1}{2 \eta}, \label{bottom and top of bands of staggered DMI-AFM configuration}
\end{equation}
where $\pm$ signs correspond to $\omega_n^b$ and $\omega_p^t$ respectively. Then, the gap between nutational and precessional bands is $1/\eta$, which is suppressed with increasing $\eta$.

\begin{table*}[t!]
\caption{Amplitudes, phases, and ellipticity angles of the SWs on top of a canted configuration for staggered DMI.}
\begin{tblr}{
  colspec = {*{13}{c}},
  rows = {valign = m}
}
\hline \hline
& & $\omega_n^+$ & $- \omega_n^+$ & & $\omega_n^-$ & $- \omega_n^-$ & & $\omega_p^+$ & $- \omega_p^+$ & & $\omega_p^-$ & $- \omega_p^-$ \\
\hline
$A_{\theta, 2 j}$ & & $\rho_n^+$ & $\rho_n^+$ & & $\rho_n^-$ & $\rho_n^-$ & & $\rho_p^+$ & $\rho_p^+$ & & $\rho_p^-$ & $\rho_p^-$ \\
$A_{\phi, 2 j}$ & & $\frac{1}{\omega_n^+}$ & $\frac{1}{\omega_n^+}$ & & $\frac{1}{\omega_n^-}$ & $\frac{1}{\omega_n^-}$ & & $\frac{1}{\omega_p^+}$ & $\frac{1}{\omega_p^+}$ & & $\frac{1}{\omega_p^-}$ & $\frac{1}{\omega_p^-}$ \\
$\delta_{\theta,2 j}$ & & 0 & 0 & & $\pi$ & $\pi$ & & $\pi$ & $\pi$ & & 0 & 0 \\
$\delta_{\phi,2 j}$ & & $- \frac{\pi}{2}$ & $\frac{\pi}{2}$ & & $\frac{\pi}{2}$ & $- \frac{\pi}{2}$ & & $- \frac{\pi}{2}$ & $\frac{\pi}{2}$ & & $\frac{\pi}{2}$ & $- \frac{\pi}{2}$ \\
\hline
$A_{\theta, 2 j + 1}$ & & $\rho_n^+$ & $\rho_n^+$ & & $\rho_n^-$ & $\rho_n^-$ & & $\rho_p^+$ & $\rho_p^+$ & & $\rho_p^-$ & $\rho_p^-$ \\
$A_{\phi, 2 j + 1}$ & & $\frac{1}{\omega_n^+}$ & $\frac{1}{\omega_n^+}$ & & $\frac{1}{\omega_n^-}$ & $\frac{1}{\omega_n^-}$ & & $\frac{1}{\omega_p^+}$ & $\frac{1}{\omega_p^+}$ & & $\frac{1}{\omega_p^-}$ & $\frac{1}{\omega_p^-}$ \\
$\delta_{\theta,2 j + 1}$ & & $\pi$ & $\pi$ & & $\pi$ & $\pi$ & & 0 & 0 & & 0 & 0 \\
$\delta_{\phi,2 j + 1}$ & & $\frac{\pi}{2}$ & $- \frac{\pi}{2}$ & & $\frac{\pi}{2}$ & $- \frac{\pi}{2}$ & & $\frac{\pi}{2}$ & $- \frac{\pi}{2}$ & & $\frac{\pi}{2}$ & $- \frac{\pi}{2}$ \\
\hline
$\chi_{2 j}$ & & $- X_n^+$ & $- X_n^+$ & & $- X_n^-$ & $- X_n^-$ & & $X_p^+$ & $X_p^+$ & & $X_p^-$ & $X_p^-$ \\
$\chi_{2 j + 1}$ & & $- X_n^+$ & $- X_n^+$ & & $- X_n^-$ & $- X_n^-$ & & $X_p^+$ & $X_p^+$ & & $X_p^-$ & $X_p^-$ \\ \hline \hline
\end{tblr}
\label{spin-wave parameters of canted configuration}
\end{table*}

\vspace{-1em}

\subsubsection{Sublattice amplitude ratio}

The amplitude ratio of SWs between two sublattices can be represented by $\vert \rho_{n,p} \vert$, which varies with $k$, as displayed in Fig. \ref{dispersion and amplitude ratio of staggered DMI-AFM configuration}(b). At $k = 0$, the sublattice amplitude ratios reach their maximum values. In particular, the ratio is close to $1$ for the precessional mode. Under the conditions $\omega_K \ll \omega_E$ and $\omega_D \ll \omega_E$, $\vert \rho_p \vert$ reduces to
\begin{eqnarray}
\rho_p^0 & \approx & 1 - \frac{1}{2} \frac{1}{1 + 4 \eta \omega_E} \frac{\omega_D^2}{\omega_E^2} - \sqrt{2} \frac{1}{\sqrt{1 + 4 \eta \omega_E}} \times \notag \\ && \left( 1 - \frac{\omega_D^2}{4 \omega_E \omega_K} \right) \sqrt{\frac{\omega_K}{\omega_E}}, \label{rho for precessional mode at k = 0}
\end{eqnarray}
which remains close to unity because both $\omega_K$ and $\omega_D$ are small compared with $\omega_E$. This indicates that the amplitudes of precessional modes in two sublattices, in the long-wavelength limit, are almost identical. With increasing $k$ from $k = 0$, $\vert \rho_{n,p} \vert$ decreases gradually and vanishes at $k = \pi/(2 a)$. This implies that, for short-wavelength SWs with $\lambda = 4 a$, the excitation is localized on a single sublattice. In addition, the sublattice amplitude ratios of mode I and mode II are mutually reciprocal, as clearly seen from the $A_\theta$ and $A_\phi$ data presented in Tab. \ref{spin-wave parameters of staggered DMI-AFM configuration}. This feature is also illustrated schematically in Tab. \ref{schematic modes of staggered DMI-AFM configuration}.

\vspace{-1em}

\subsubsection{Chirality and phase}

Since we adopt local coordinate frames (shown in Tab. \ref{schematic modes of staggered DMI-AFM configuration}) throughout spin-wave calculations, chirality is defined as the rotational sense about the equilibrium magnetization of each individual sublattice \cite{Z. Q. Qiu}. Following this definition, modes I and II exhibit opposite signs of $\chi$ with $\vert \chi \vert = \pi/4$, indicating opposite chiralities that correspond to left-handed and right-handed circular polarizations, respectively. Moreover, the corresponding chirality exhibits an opposite orientation between two sublattices. By contrast, some previous studies choose the magnetic field direction, aligned with the static magnetization of one sublattice, as the rotational axis. With this alternative convention, the two sublattices have the same chirality.

Here, flipping the sign of frequency can not reverse the chirality of SWs. As seen by comparing the phase characteristics of the $\omega_{n,p}$-branch with those of the corresponding $-\omega_{n,p}$-branch in Table \ref{spin-wave parameters of staggered DMI-AFM configuration}, the phase differences between the $\mathbf{e}_\theta$ and $\mathbf{e}_\phi$ components of the spin-wave fluctuations have opposite signs. According to the definition of ellipticity angle Eq. (\ref{ellipticity angle}), the chirality is not altered when combining with the sign reversal of $\omega$.

Additionally, in Tab. \ref{schematic modes of staggered DMI-AFM configuration}, we schematically illustrate the phases of all modes at a specific lattice site at a given instant in time. It reveals that, for $0 < k a < \pi/2$, the two sublattices oscillate with a phase difference of $\varphi_0$ for nutation and with a phase difference of $\pi + \varphi_0$ for precession; for $\pi/2 < k a < \pi$, the phase relations are reversed. Without DMI, $\varphi_0$ is zero.

\vspace{-1em}

\subsection{Spin waves on top of canted configuration}

When $\omega_D > \omega_D^c$, the canted configuration is preferred in equilibrium. After calculating the spin-wave modes, as detailed in Appendix \ref{Spin-wave solution in canted configuration}, the components of the spin-wave fluctuations can be expressed in the form of Eqs. (\ref{components of spin-wave fluctuation for 2j sublattice}) and (\ref{components of spin-wave fluctuation for 2j+1 sublattice}). In this case, the spectrum is nondegenerate. Eight spin-wave branches are obtained, four with positive frequencies and four corresponding negative-frequency branches. Their amplitudes and initial phases, as well as the ellipticity angles [calculated from Eq. (\ref{ellipticity angle})] are listed in Tab. \ref{spin-wave parameters of canted configuration} for every branches of SWs. Additionally, we plot the dispersions and the dependence of ellipticity angles on the wave number in Fig. \ref{dispersion and ellipticity angle of canted configuration}.

In Tab \ref{spin-wave parameters of canted configuration}, the eigenfrequencies of nutational and precessional SWs read
\begin{equation}
\omega_{n,p}^\pm (k) = \frac{\sqrt{1 + \eta \Gamma_\pm \pm \sqrt{\left( 1 + \eta \Gamma_\pm \right)^2 - \eta^2 \Lambda_\pm^2}}}{\sqrt{2} \eta}, \label{omega of canted configuration}
\end{equation}
where $\Gamma_\pm = 2 ( f_\pm + g_\pm )$, and $\Lambda_\pm = 4 \sqrt{f_\pm g_\pm}$, with $f_\pm(k) = \sqrt{\omega_E^2 + \omega_D^2} - \sqrt{\omega_E^2 + \omega_D^{c 2}} + \omega_E ( 1 \mp \cos k a )$, and $g_\pm(k) = \sqrt{\omega_E^2 + \omega_D^2} ( 1 \pm \cos k a )$. In Eq. (\ref{omega of canted configuration}), the subscripts ${n,p}$ refer, respectively, to the upper and lower signs of $\pm$ in front of the square root. It's worth noting that $f_\pm(k)$ remains positive for any $k$ if $\omega_D > \omega_D^c$. By contrast, when $\omega_D < \omega_D^c$, $f_\pm(k)$ becomes negative for some values of $k$, yielding $\Lambda^2_\pm < 0$. This in turn implies that $\omega_p^\pm$ is complex according to Eq. (\ref{omega of canted configuration}), signifying the instability of canted configuration.

In Tab \ref{spin-wave parameters of canted configuration}, the parameters $\rho_{n,p}^\pm$ are given by
\begin{eqnarray}
\rho_{n,p}^\pm = \pm \frac{1}{\eta \left( \omega^\pm_{n,p} \right)^2 - 2 f_\pm}, \label{rho of canted configuration}
\end{eqnarray}
with the upper and lower signs in front of the fraction corresponding to the subscripts $n$ and $p$ of $\rho_{n,p}^\pm$, respectively. It can be verified that $\rho_{n,p}^\pm > 0$.

\vspace{-1em}

\subsubsection{Long-wavelength regime}

\begin{figure*}[t!]
    \centering
    \begin{minipage}[b]{0.49\textwidth}
        \centering
        \includegraphics[clip, trim=0cm 2cm 1cm 1cm, width=\textwidth]{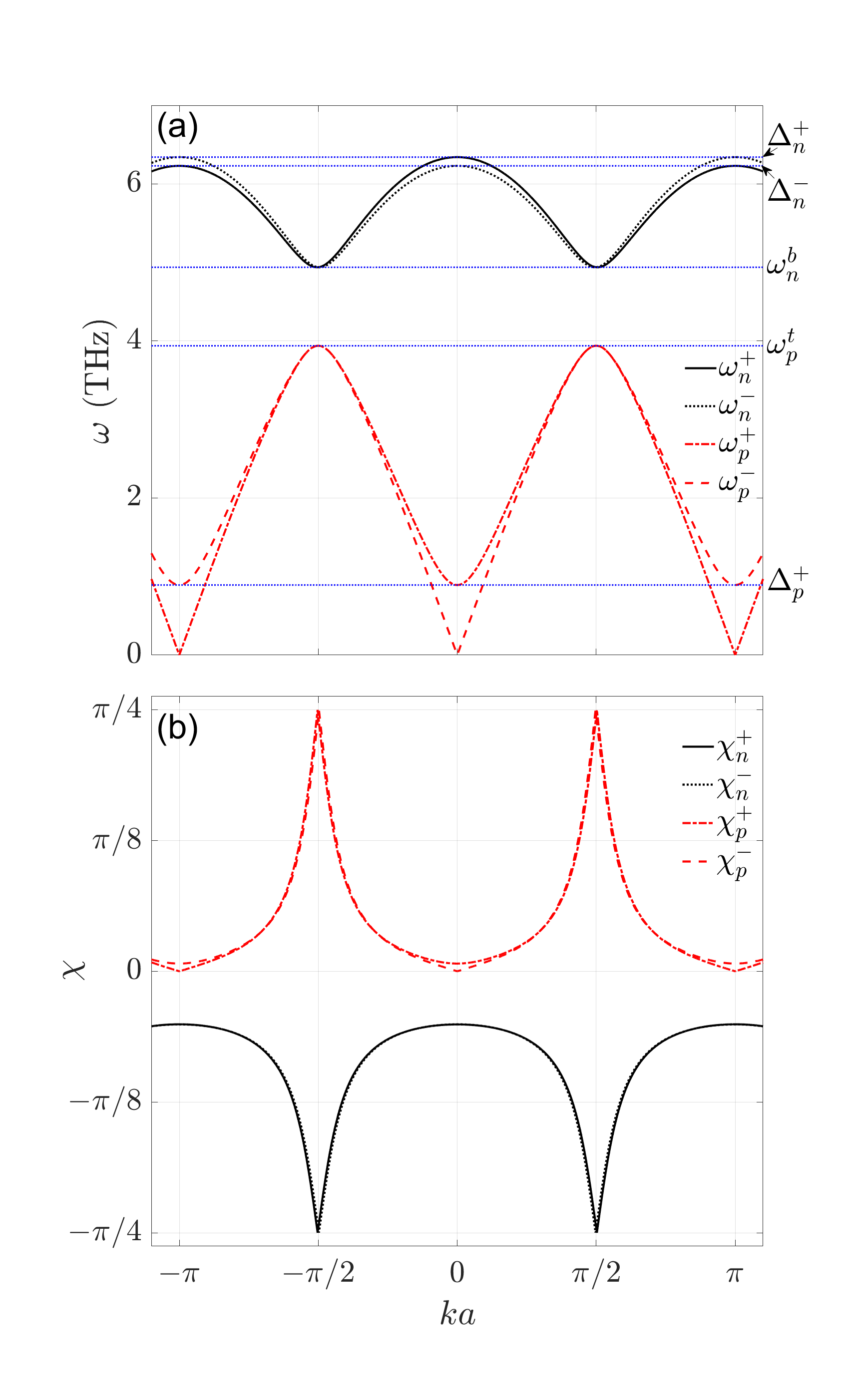}
    \end{minipage}
    \hspace{0.001\textwidth}%
    \begin{minipage}[b]{0.49\textwidth}
        \centering
        \includegraphics[clip, trim=1cm 2cm 0cm 1cm, width=\textwidth]{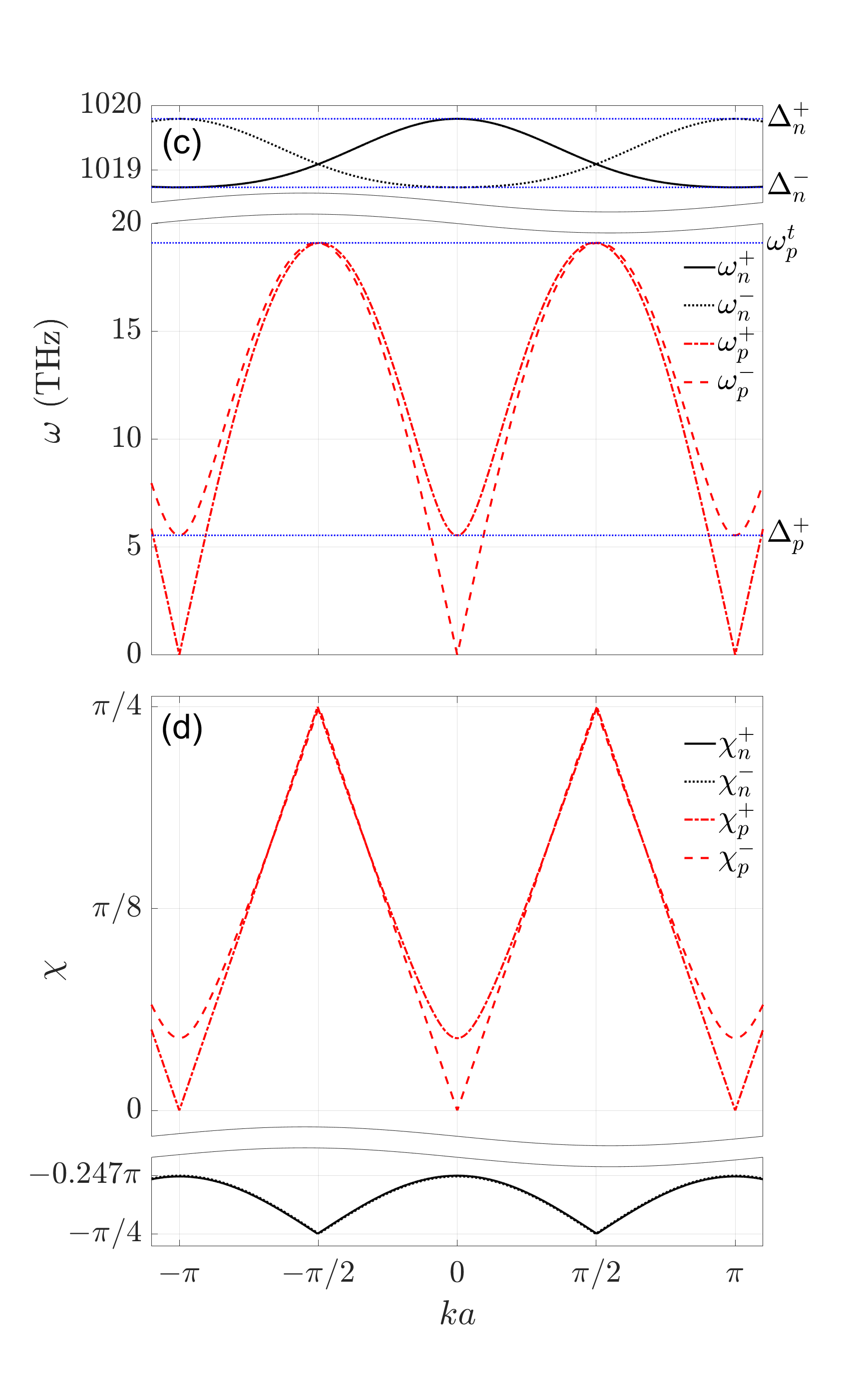}
    \end{minipage}
    \caption{Dispersions (a), (c) and ellipticity angles (b), (d) of the SWs on top of a canted configuration for the case of staggered DMI. (a) and (b) correspond to the regime $\eta > \eta_c$, while (c) and (d) are obtained for $\eta < \eta_c$. In (b) and (d), $\chi^\pm_n = - X^\pm_n$ and $\chi^\pm_p = X^\pm_p$. Here, $\eta = 1$ ps for (a) and (b), and $\eta = 1$ fs for (c) and (d). The DMI strength is set to $\omega_D = 2 \omega_D^c$. The other parameters follow those adopted in Fig. \ref{dispersion and amplitude ratio of staggered DMI-AFM configuration}.} \label{dispersion and ellipticity angle of canted configuration}
\end{figure*}

In the long-wavelength limit, the $\omega_n^+$- and $\omega_p^+$-branches can be described by
\begin{equation}
\omega^+_{n,p} = \sqrt{\Delta^{+ 2}_{n,p} \mp v^{+ 2}_{n,p} k^2}, \label{long-wavelength dispersion of canted configuration}
\end{equation}
with the upper and lower signs of $\mp$ corresponding to the $n$ and $p$ branches, respectively. In Eq. (\ref{long-wavelength dispersion of canted configuration}), the uniform precession frequency
\begin{equation}
\Delta^+_{n,p} = \frac{\sqrt{1 + 2 \eta \xi \pm \sqrt{\left( 1 + 2 \eta \xi \right)^2 - 32 \eta^2 \widetilde{\omega}_E \xi_-}}}{\sqrt{2} \eta}, \label{omega of upper branches at k = 0}
\end{equation}
and the characteristic velocity
\begin{equation}
\! v^+_{n,p} = a \! \sqrt{ \! \frac{\zeta_- \! \left[ \! 1 \pm \! \sqrt{ \! \left( 1 + 2 \eta \xi \right)^2 \! - 32 \eta^2 \widetilde{\omega}_E \xi_-} \right] \! + 2 \eta \zeta_+ \xi_+}{2 \eta \sqrt{\left( 1 + 2 \eta \xi \right)^2 - 32 \eta^2 \widetilde{\omega}_E \xi_-}}}, \label{velocity of upper branches at k = 0}
\end{equation}
with $\widetilde{\omega}_E = \sqrt{\omega_E^2 + \omega_D^2}$, $\xi = 3 \sqrt{\omega_E^2 + \omega_D^2} - \sqrt{\omega_E^2 + \omega_D^{c 2}}$, $\xi_\pm $=$ \sqrt{\omega_E^2 + \omega_D^2} \pm \sqrt{\omega_E^2 + \omega_D^{c 2}}$, and $\zeta_\pm = \sqrt{\omega_E^2 + \omega_D^2} \pm \omega_E$.
In Eqs. (\ref{omega of upper branches at k = 0}) and (\ref{velocity of upper branches at k = 0}), the upper and lower signs in front of the square root correspond to the $n$ and $p$ branches, respectively. In this case, the group velocity and wave number have opposite signs for the nutational SWs, which is a characteristic feature of backward SWs.

The $\omega^-_p$-branch is gapless, as shown in Figs. \ref{dispersion and ellipticity angle of canted configuration}(a) and \ref{dispersion and ellipticity angle of canted configuration}(c). Its dispersion can be approximated as:
\begin{equation}
\omega^-_p = v^-_p \vert k \vert,
\end{equation}
where
\vspace{-1em}
\begin{equation}
v^-_p = a \sqrt{\frac{2 \sqrt{\omega_E^2 + \omega_D^2} \left( 2 \omega_E + \xi_- \right)}{1 + 2 \eta \left( 2 \omega_E + \xi_- \right)}}.
\end{equation}
The presence of spin inertia diminishes the velocity of long-wavelength SW. It should be mentioned that a Goldstone mode exists for the precessional SWs in the canted configuration, in contrast to the gapped SWs in the AFM configuration [see Fig. \ref{dispersion and amplitude ratio of staggered DMI-AFM configuration}(a)]. This difference arises because the canted configuration breaks the rotational symmetry about the $z$ axis, whereas the AFM configuration preserves this symmetry.

Comparison of the dotted curves in Figs. \ref{dispersion and ellipticity angle of canted configuration}(a) and \ref{dispersion and ellipticity angle of canted configuration}(c) shows that the long-wavelength dispersion of the $\omega_n^-$ branch differs substantially between the $\eta > \eta_c$ and $\eta < \eta_c$ regimes. The critical value $\eta_c$ is derived as
\begin{equation}
\eta_c = \frac{1}{2 \omega_E} \frac{\sqrt{\omega_E^2 + \omega_D^2} - \omega_E}{\sqrt{\omega_E^2 + \omega_D^2} - \sqrt{\omega_E^2 + \omega_D^{c 2}} + 2 \omega_E}.
\end{equation}
The dispersions for $\eta > \eta_c$ and $\eta < \eta_c$ can be written as a single equation,
\begin{equation}
\omega^-_n = \sqrt{\left( \Delta^-_n \right)^2 - \sgn \left( 1 - \frac{\eta_c}{\eta} \right) \left( v^-_n \right)^2 k^2},
\end{equation}
where
\begin{equation}
\Delta_n^- = \frac{\sqrt{1 + 2 \eta \left( 2 \omega_E + \xi_- \right)}}{\eta},
\end{equation}
\vspace{-1em}
and
\begin{equation}
v^-_n = a \sqrt{\frac{2 \omega_E \left( 2 \omega_E + \xi_- \right)}{1 + 2 \eta \left( 2 \omega_E + \xi_- \right)} \left \vert 1 - \frac{\eta_c}{\eta} \right \vert}.
\end{equation}
The velocity $v_n^-$ becomes zero at $\eta = \eta_c$. Accordingly, the nutational dispersion is nearly flat in the long-wavelength regime, giving rise to a sharp peak for the density of states, and yielding strong and narrow signals for the specific heat and neutron scattering spectra.

Fig. \ref{dispersion and ellipticity angle of canted configuration}(c) also indicates that the two nutational branches exhibit opposite dispersion behaviors for $\eta < \eta_c$: one reaches its maximum frequency at $k = 0$ (with minima at $k = \pm \pi/a$), while the other shows the opposite trend, attaining its minimum frequency at $k = 0$ and maxima at $k = \pm \pi/a$.

\vspace{-1em}

\subsubsection{Short-wavelength regime}

\begin{figure}[t]
\includegraphics[clip, trim = 0cm 22cm 0cm 3cm, width=0.5\textwidth]{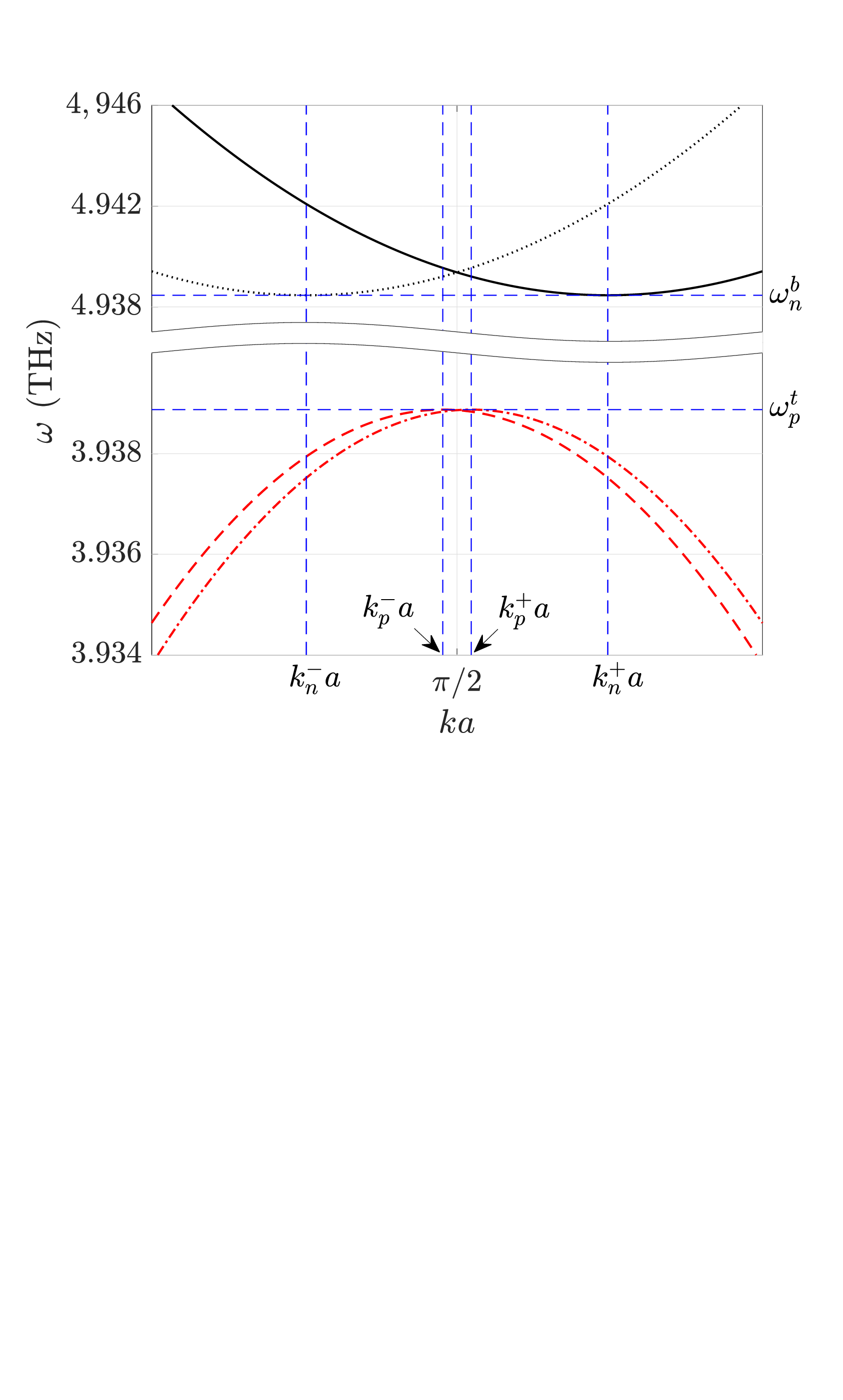}
\caption{(color online). Enlarged view of Figs. \ref{dispersion and ellipticity angle of canted configuration}(a) around $k = \pi/(2 a)$.}
\label{dispersion and ellipticity angle of canted configuration zoomin}
\end{figure}

It is interesting to examine the behaviors of SWs near $k = \pm \pi/(2 a)$, as highlighted in the magnified plot of Fig. \ref{dispersion and ellipticity angle of canted configuration zoomin}. At $k = \pm \pi/(2 a)$, frequency degeneracy emerges separately for the two precessional branches and the two nutational branches, marking the unique degeneracy position for both types of excitations.  At $k = k_p^\pm = (\pi/2 \pm \varphi_p)/a$, both precessional dispersion curves attain their maximum frequencies, with identical upper cutoff frequencies for the two branches. Similarly, the minimal frequencies of the two nutational dispersion curves are reached at $k = k_n^\pm = (\pi/2 \pm \varphi_n)/a$, again with identical cutoff frequencies. The values of $\varphi_{n,p}$ are given by
\begin{eqnarray}
\varphi_{n,p} &=& \arcsin \left \{ \frac{1}{\zeta_+} \left[ \omega_K \pm \frac{1}{2 \eta} \frac{\zeta_-}{\zeta_+} \times \right. \right. \notag \\ && \left. \left. \frac{\sqrt{\omega_E + 2 \eta \zeta_+ \left( 2 \omega_E + \xi_- \right)} \pm \sqrt{\omega_E}}{\sqrt{\omega_E}} \right] \right \},
\end{eqnarray}
where the subscripts $n$ and $p$ correspond to the upper and lower signs of $\pm$, respectively. The quantities $\zeta_\pm$ and $\xi_-$ are defined below Eq. (\ref{velocity of upper branches at k = 0}). For typical experimental parameters, $\varphi_{n,p}$ is small, making it difficult to distinguish $k_{n,p}^+$ and $k_{n,p}^-$ in Fig. \ref{dispersion and ellipticity angle of canted configuration}. To resolve this small separation, an enlarged view is shown in Fig. \ref{dispersion and ellipticity angle of canted configuration zoomin}. The bottom of the nutational dispersion and the top of the precessional dispersion are expressed as
\begin{equation}
\omega_{n(p)}^{b(t)} = \sqrt{\widetilde{\omega}_E} \frac{\sqrt{\omega_E + 2 \eta \zeta_+ \left( 2 \omega_E + \xi_- \right)} \pm \sqrt{\omega_E}}{\eta \zeta_+}, \label{bottom of nutation and top of precession of canted configuration}
\end{equation}
where the upper and lower signs of $\pm$ before the square correspond to $\omega_n^b$ and $\omega_p^t$ respectively, and $\widetilde{\omega}_E$, $\zeta_+$ and $\xi_-$ are defined below Eq. (\ref{velocity of upper branches at k = 0}). From Eq. (\ref{bottom of nutation and top of precession of canted configuration}), we get the gap between nutational and precessional bands,
\begin{eqnarray}
\Delta_{n-p} &=& \frac{1}{\eta} \frac{2 \sqrt{\omega_E} \left( \omega_E^2 + \omega_D^2 \right)^{\frac{1}{4}}}{\sqrt{\omega_E^2 + \omega_D^2} + \omega_E}, \notag \\
&=& \frac{1}{\eta} \left[ 1 - \frac{1}{32} \frac{\omega_D^4}{\omega_E^4} + O \left( \frac{\omega_D^6}{\omega_E^6} \right) \right].
\end{eqnarray}
For the case that $\omega_D \ll \omega_E$, the gap is about $1/\eta$. It should be noted that, for $\eta < \eta_c$, the nutational branches exhibit no extrema at $k = \pm k_n^\pm$, as seen in Fig. \ref{dispersion and ellipticity angle of canted configuration}(c).

\vspace{-1em}

\subsubsection{Chirality and polarization}

We now turn to the chirality and polarization, both characterized by a single physical quantity: the ellipticity angle $\chi$ defined in Eq. (\ref{ellipticity angle}). Here, we find that the two sublattices have the same ellipticity angle, i.e. $\chi_n^\pm = - X_n^\pm$ and $\chi_p^\pm = X_p^\pm$, where
\begin{equation}
X_{n,p}^\pm = \frac{1}{2} \arcsin \left( \frac{2 \rho_{n,p}^\pm \omega_{n,p}^\pm}{1 + \rho_{n,p}^{\pm 2} \omega_{n,p}^{\pm 2}} \right).
\end{equation}
The wave number dependence of $\chi$ for nutational and precessional modes is presented in Figs. \ref{dispersion and ellipticity angle of canted configuration}(b) and \ref{dispersion and ellipticity angle of canted configuration}(d).

In contrast to SWs in conventional AFM configurations, the nutational modes here satisfy $\chi_n^\pm < 0$, while the precessional modes are characterized by $\chi_p^\pm > 0$. This indicates that all nutational branches exhibit left-handed chirality, whereas all precessional branches are right-handed, as also summarized in Tab. \ref{spin-wave parameters of canted configuration}. Another observation is that the chirality is the same for the positive- and negative-frequency branches, as summarized in Tab. \ref{spin-wave parameters of canted configuration}. This result indicates that the chirality depends not only on the sign of frequency, but also on the phase difference between two components of spin-wave fluctuation. It can be inferred from Tab. \ref{spin-wave parameters of canted configuration} that the phase difference ($\delta_\phi - \delta_\theta$) is opposite for the positive- and negative-frequency branches. Thus, according to the definition of ellipticity angle in Eq. (\ref{ellipticity angle}), $\chi$ is unchanged when flipping the sign of frequency. Finally, because the phase difference is identical for two sublattices, i.e. $\delta_{\phi,2 j + 1} - \delta_{\theta,2 j + 1} = \delta_{\phi,2 j} - \delta_{\theta,2 j}$, the chirality is the same for two sublattices, as seen by comparing $\chi_{2 j}$ and $\chi_{2 j + 1}$ in Tab. \ref{spin-wave parameters of canted configuration}.

Polarization exhibits a strong dependence on the wave number. For the precessional $\omega_p^-$ branch, shown by the dashed curves in Figs. \ref{dispersion and ellipticity angle of canted configuration}(b) and \ref{dispersion and ellipticity angle of canted configuration}(d), one finds $\chi_p^- = 0$ at $k = 0$, corresponding to linear polarization. As $k$ increases, $\chi$ rises continuously, indicating that the mode evolves from elliptical polarization toward circular polarization. When $k$ reaches $k_p^-$, the ellipticity angle becomes
\begin{eqnarray}
\chi_p^- \left( k = k_p^- \right) &=& \frac{1}{2} \arcsin \left[ \frac{2 \sqrt{\omega_E} \left( \omega_E^2 + \omega_D^2 \right)^{\frac{1}{4}}}{\sqrt{\omega_E^2 + \omega_D^2} + \omega_E} \right] \notag \\
&=& \frac{\pi}{4} - \frac{\omega_D^2}{8 \omega_E^2} + O \left( \frac{\omega_D^4}{\omega_E^4} \right).
\end{eqnarray}
Given the condition $\omega_D \ll \omega_E$, $\chi_p^-$ at $k = k_p^-$ is very close to $\pi/4$ and the corresponding mode is nearly circularly polarized. For the precessional $\omega_p^+$ branch, the polarization behavior follows a similar trend, with the only difference being $\chi_p^+ \neq 0$ at $k = 0$. With $k$ increasing from $0$, $\chi_p^+$ rises gradually and eventually approaches $\pi/4$ at $k = k_p^+$. The two branches share the same maximum ellipticity angle, i.e. $\chi_p^+ \left( k = k_p^+ \right) = \chi_p^- \left( k = k_p^- \right)$.

For nutational SWs, $\vert \chi_n^\pm \vert$ attains its minimum at $k = 0$ and its maximum at $k = k_n^\pm$, satisfying $\chi_n^+ \left( k = k_n^+ \right) = \chi_n^- \left( k = k_n^- \right) = - \chi_p^- \left( k = k_p^- \right)$. Furthermore, the $\chi$-$k$ curves of the two nutational branches are very close to each other and hardly distinguishable, as displayed by the solid and dotted curves in Figs. \ref{dispersion and ellipticity angle of canted configuration}(b) and \ref{dispersion and ellipticity angle of canted configuration}(d). For both nutational branches, the ellipticity (defined by $\vert \chi_n^\pm \vert$) increases monotonically with $k$ increasing from $0$ and approaches $\pi/4$ at $k = k_n^+$ and $k = k_n^-$ respectively. In summary, precessional SWs exhibit nearly linear polarization in the long-wavelength limit. Near the boundary of Brillouin zone [$k = \pm \pi/(2 a)$], both precessional and nutational SWs approach nearly circular polarization.

\vspace{-1em}

\section{Inertial spin waves in AFM with homogeneous DMI} \label{Inertial SWs in AFM with homogeneous DMI}

\subsection{Spin waves on top of AFM configuration}

\begin{table*}[t]
\caption{Amplitudes, phases, and ellipticity angles of the SWs on top of an AFM configuration for homogeneous DMI. Here, $\text{H}(\lambda^\pm)$ is the unit step function and $\lambda^\pm = \sgn \left[ \cos \left( k a \mp \varphi_d \right) \right]$.}
\begin{tblr}{
  colspec = {*{13}{c}},
  rows = {valign = m}
}
\hline \hline
& & $\omega_n^+$ & $- \omega_n^+$ & & $\omega_n^-$ & $- \omega_n^-$ & & $\omega_p^+$ & $- \omega_p^+$ & & $\omega_p^-$ & $- \omega_p^-$ \\
\hline
$A_{\theta, 2 j}$ & & $\frac{\left \vert \rho_n^+ \right \vert}{\omega_n^+}$ & $\frac{1}{\omega_n^+}$ & & $\frac{1}{\omega_n^-}$ & $\frac{\left \vert \rho_n^- \right \vert}{\omega_n^-}$ & & $\frac{\left \vert \rho_p^+ \right \vert}{\omega_p^+}$ & $\frac{1}{\omega_p^+}$ & & $\frac{1}{\omega_p^-}$ & $\frac{\left \vert \rho_p^- \right \vert}{\omega_p^-}$ \\
$A_{\phi, 2 j}$ & & $\frac{\left \vert \rho_n^+ \right \vert}{\omega_n^+}$ & $\frac{1}{\omega_n^+}$ & & $\frac{1}{\omega_n^-}$ & $\frac{\left \vert \rho_n^- \right \vert}{\omega_n^-}$ & & $\frac{\left \vert \rho_p^+ \right \vert}{\omega_p^+}$ & $\frac{1}{\omega_p^+}$ & & $\frac{1}{\omega_p^-}$ &  $\frac{\left \vert \rho_p^- \right \vert}{\omega_p^-}$ \\
$\delta_{\theta,2 j}$ & & {\footnotesize $\pi \text{H} (- \lambda^+)$} & {\footnotesize $\pi \text{H} (\lambda^+)$} & & {\footnotesize $\pi \text{H} (\lambda^-)$} & {\footnotesize $\pi \text{H} (- \lambda^-)$} & & {\footnotesize $\pi \text{H} (\lambda^+)$} & {\footnotesize $\pi \text{H} (- \lambda^+)$} & & {\footnotesize $\pi \text{H} (- \lambda^-)$} & {\footnotesize $\pi \text{H} (\lambda^-)$} \\
$\delta_{\phi,2 j}$ & & {\footnotesize $\lambda^+ \frac{\pi}{2}$} & {\footnotesize $- \lambda^+ \frac{\pi}{2}$} & & {\footnotesize $\lambda^- \frac{\pi}{2}$} & {\footnotesize $- \lambda^- \frac{\pi}{2}$} & & {\footnotesize $- \lambda^+ \frac{\pi}{2}$} & {\footnotesize $\lambda^+ \frac{\pi}{2}$} & & {\footnotesize $- \lambda^- \frac{\pi}{2}$} & {\footnotesize $\lambda^- \frac{\pi}{2}$} \\
\hline
$A_{\theta, 2 j + 1}$ & & $\frac{1}{\omega_n^+}$ & $\frac{\left \vert \rho_n^+ \right \vert}{\omega_n^+}$ & & $\frac{\left \vert \rho_n^- \right \vert}{\omega_n^-}$ & $\frac{1}{\omega_n^-}$ & & $\frac{1}{\omega_p^+}$ & $\frac{\left \vert \rho_p^+ \right \vert}{\omega_p^+}$ & & $\frac{\left \vert \rho_p^- \right \vert}{\omega_p^-}$ & $\frac{1}{\omega_p^-}$ \\
$A_{\phi, 2 j + 1}$ & & $\frac{1}{\omega_n^+}$ & $\frac{\left \vert \rho_n^+ \right \vert}{\omega_n^+}$ & & $\frac{\left \vert \rho_n^- \right \vert}{\omega_n^-}$ & $\frac{1}{\omega_n^-}$ & & $\frac{1}{\omega_p^+}$ & $\frac{\left \vert \rho_p^+ \right \vert}{\omega_p^+}$ & & $\frac{\left \vert \rho_p^- \right \vert}{\omega_p^-}$ & $\frac{1}{\omega_p^-}$ \\
$\delta_{\theta,2 j + 1}$ & & 0 & $\pi$ & & $\pi$ & 0 & & 0 & $\pi$ & & $\pi$ & 0 \\
$\delta_{\phi,2 j + 1}$ & & $- \frac{\pi}{2}$ & $\frac{\pi}{2}$ & & $- \frac{\pi}{2}$ & $\frac{\pi}{2}$ & & $- \frac{\pi}{2}$ & $\frac{\pi}{2}$ & & $- \frac{\pi}{2}$ & $\frac{\pi}{2}$ \\
\hline
$\chi_{2 j}$ & & $\frac{\pi}{4}$ & $- \frac{\pi}{4}$ & & $- \frac{\pi}{4}$ & $\frac{\pi}{4}$ & & $\frac{\pi}{4}$ & $- \frac{\pi}{4}$ & & $- \frac{\pi}{4}$ & $\frac{\pi}{4}$ \\
$\chi_{2 j + 1}$ & & $- \frac{\pi}{4}$ & $\frac{\pi}{4}$ & & $\frac{\pi}{4}$ & $- \frac{\pi}{4}$ & & $- \frac{\pi}{4}$ & $\frac{\pi}{4}$ & & $\frac{\pi}{4}$ & $- \frac{\pi}{4}$ \\
\hline \hline
\end{tblr}
\label{spin-wave parameters of homogeneous DMI-AFM configuration}
\end{table*}

\begin{table*}[t]
\centering
\caption{Schematic diagrams of the SWs on top of an AFM configuration for the case of homogeneous DMI. The blue solid (red dotted) circles denote the trajectories of the tips of $\mathbf{m}_{ 2 j}$ ($\mathbf{m}_{ 2 j + 1}$). The arrows on the blue solid (red dotted) circles signify the rotational senses of $\mathbf{m}_{ 2 j}$ ($\mathbf{m}_{ 2 j + 1}$) around  $- \mathbf{r}_r^{ 2 j}$ ($\mathbf{e}_r^{ 2 j + 1}$). The blue solid (red dotted) vectors line represent the fluctuation $\delta \mathbf{m}_{2 j}$ ($\delta \mathbf{m}_{2 j + 1}$).}
\begin{tblr}{
  colspec = {*{12}{c}},
  rows = {valign = m}
}
\hline \hline
& $\omega_n^+$ & $- \omega_n^+$ & & $\omega_n^-$ & $- \omega_n^-$ & & $\omega_p^+$ & $- \omega_p^+$ & & $\omega_p^-$ & $- \omega_p^-$ \\
\hline
$\lambda_\pm > 0$ & \adjustbox{valign=m}{\includegraphics[clip, trim = 0cm 0cm 0cm 0cm, width=0.07\textwidth]{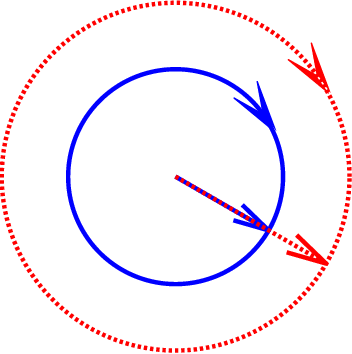}} & \adjustbox{valign=m}{\includegraphics[clip, trim = 0cm 0cm 0cm 0cm, width=0.07\textwidth]{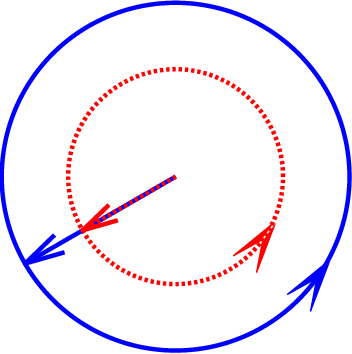}} & & \adjustbox{valign=m}{\includegraphics[clip, trim = 0cm 0cm 0cm 0cm, width=0.07\textwidth]{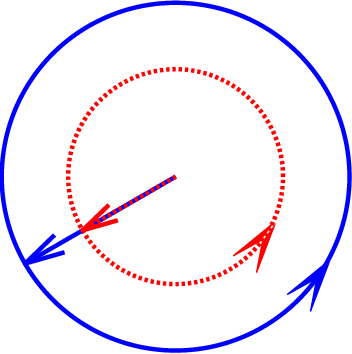}} & \adjustbox{valign=m}{\includegraphics[clip, trim = 0cm 0cm 0cm 0cm, width=0.07\textwidth]{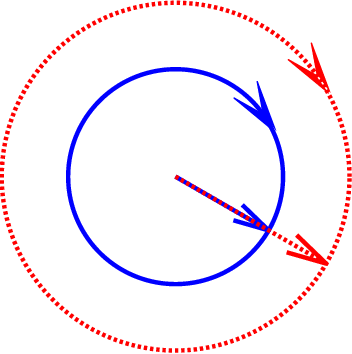}} & & \adjustbox{valign=m}{\includegraphics[clip, trim = 0cm 0cm 0cm 0cm, width=0.07\textwidth]{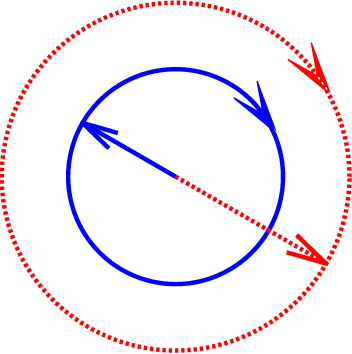}} & \adjustbox{valign=m}{\includegraphics[clip, trim = 0cm 0cm 0cm 0cm, width=0.07\textwidth]{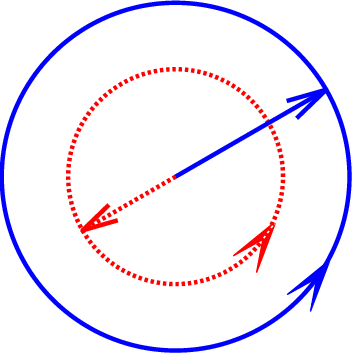}} & & \adjustbox{valign=m}{\includegraphics[clip, trim = 0cm 0cm 0cm 0cm, width=0.07\textwidth]{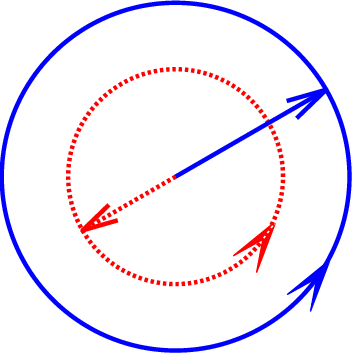}} & \adjustbox{valign=m}{\includegraphics[clip, trim = 0cm 0cm 0cm 0cm, width=0.07\textwidth]{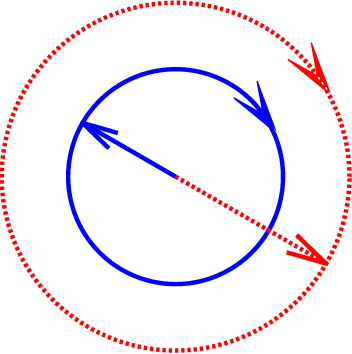}} \\
$\lambda_\pm < 0$ & \adjustbox{valign=m}{\includegraphics[clip, trim = 0cm 0cm 0cm 0cm, width=0.07\textwidth]{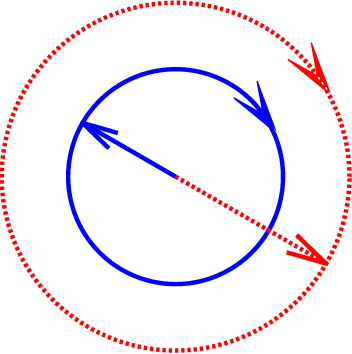}} & \adjustbox{valign=m}{\includegraphics[clip, trim = 0cm 0cm 0cm 0cm, width=0.07\textwidth]{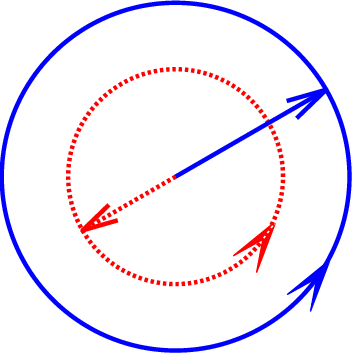}} & & \adjustbox{valign=m}{\includegraphics[clip, trim = 0cm 0cm 0cm 0cm, width=0.07\textwidth]{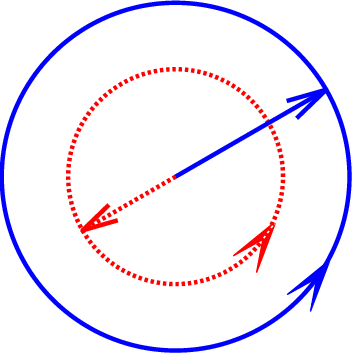}} & \adjustbox{valign=m}{\includegraphics[clip, trim = 0cm 0cm 0cm 0cm, width=0.07\textwidth]{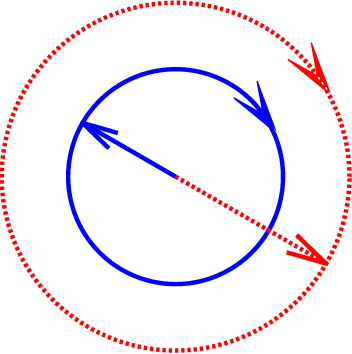}} & & \adjustbox{valign=m}{\includegraphics[clip, trim = 0cm 0cm 0cm 0cm, width=0.07\textwidth]{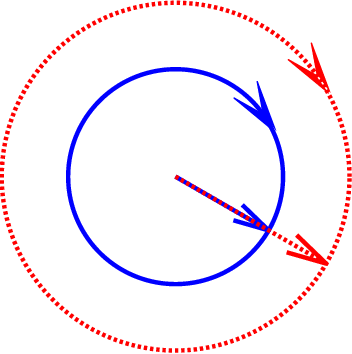}} & \adjustbox{valign=m}{\includegraphics[clip, trim = 0cm 0cm 0cm 0cm, width=0.07\textwidth]{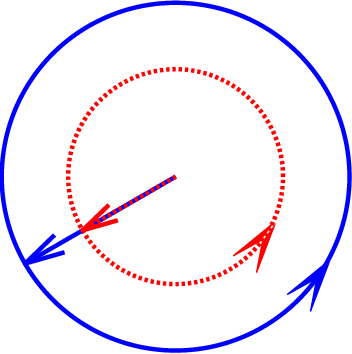}} & & \adjustbox{valign=m}{\includegraphics[clip, trim = 0cm 0cm 0cm 0cm, width=0.07\textwidth]{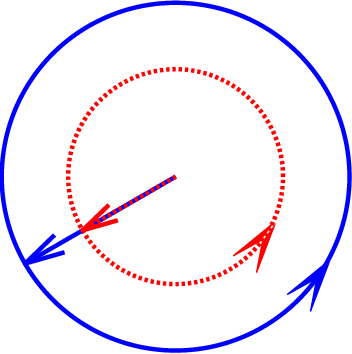}} & \adjustbox{valign=m}{\includegraphics[clip, trim = 0cm 0cm 0cm 0cm, width=0.07\textwidth]{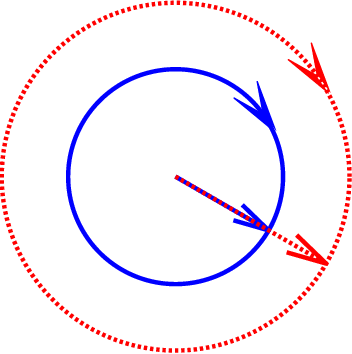}} \\
$\begin{array}{c} \lambda_+ > 0 \\ \lambda_- < 0 \end{array}$ & \adjustbox{valign=m}{\includegraphics[clip, trim = 0cm 0cm 0cm 0cm, width=0.07\textwidth]{modes_nutation_for_homo_DMI_11.eps}} & \adjustbox{valign=m}{\includegraphics[clip, trim = 0cm 0cm 0cm 0cm, width=0.07\textwidth]{modes_nutation_for_homo_DMI_12.eps}} & & \adjustbox{valign=m}{\includegraphics[clip, trim = 0cm 0cm 0cm 0cm, width=0.07\textwidth]{modes_nutation_for_homo_DMI_23.eps}} & \adjustbox{valign=m}{\includegraphics[clip, trim = 0cm 0cm 0cm 0cm, width=0.07\textwidth]{modes_nutation_for_homo_DMI_24.eps}} & & \adjustbox{valign=m}{\includegraphics[clip, trim = 0cm 0cm 0cm 0cm, width=0.07\textwidth]{modes_precession_for_homo_DMI_11.eps}} & \adjustbox{valign=m}{\includegraphics[clip, trim = 0cm 0cm 0cm 0cm, width=0.07\textwidth]{modes_precession_for_homo_DMI_12.eps}} & & \adjustbox{valign=m}{\includegraphics[clip, trim = 0cm 0cm 0cm 0cm, width=0.07\textwidth]{modes_precession_for_homo_DMI_23.eps}} & \adjustbox{valign=m}{\includegraphics[clip, trim = 0cm 0cm 0cm 0cm, width=0.07\textwidth]{modes_precession_for_homo_DMI_24.eps}} \\
$\begin{array}{c} \lambda_+ < 0 \\ \lambda_- > 0 \end{array}$ & \adjustbox{valign=m}{\includegraphics[clip, trim = 0cm 0cm 0cm 0cm, width=0.07\textwidth]{modes_nutation_for_homo_DMI_21.eps}} & \adjustbox{valign=m}{\includegraphics[clip, trim = 0cm 0cm 0cm 0cm, width=0.07\textwidth]{modes_nutation_for_homo_DMI_22.eps}} & & \adjustbox{valign=m}{\includegraphics[clip, trim = 0cm 0cm 0cm 0cm, width=0.07\textwidth]{modes_nutation_for_homo_DMI_13.eps}} & \adjustbox{valign=m}{\includegraphics[clip, trim = 0cm 0cm 0cm 0cm, width=0.07\textwidth]{modes_nutation_for_homo_DMI_14.eps}} & & \adjustbox{valign=m}{\includegraphics[clip, trim = 0cm 0cm 0cm 0cm, width=0.07\textwidth]{modes_precession_for_homo_DMI_21.eps}} & \adjustbox{valign=m}{\includegraphics[clip, trim = 0cm 0cm 0cm 0cm, width=0.07\textwidth]{modes_precession_for_homo_DMI_22.eps}} & & \adjustbox{valign=m}{\includegraphics[clip, trim = 0cm 0cm 0cm 0cm, width=0.07\textwidth]{modes_precession_for_homo_DMI_13.eps}} & \adjustbox{valign=m}{\includegraphics[clip, trim = 0cm 0cm 0cm 0cm, width=0.07\textwidth]{modes_precession_for_homo_DMI_14.eps}} \\
\hline
frame & \SetCell[c=2]{c}{\adjustbox{valign=m}{\includegraphics[clip, trim = 0cm 0cm 0cm 0cm, width=0.11\textwidth]{frame_xyz.eps}}} & & & \SetCell[c=2]{c}{\adjustbox{valign=m}{\includegraphics[clip, trim = 0cm 0cm 0cm 0cm, width=0.11\textwidth]{frame_sublattice_odd.eps}}} & & & \SetCell[c=2]{c}{\adjustbox{valign=m}{\includegraphics[clip, trim = 0cm 0cm 0cm 0cm, width=0.11\textwidth]{frame_sublattice_even.eps}}} &   \\
\hline \hline
\end{tblr}
\label{schematic of modes of homogeneous DMI-AFM configuration}
\end{table*}

Compared with Sec. \ref{secIVA}, the AFM configuration remains identical, while the DMI term is modified into the staggered form. Following the same procedure as in Sec. \ref{secIVA}, with details provided in Appendix \ref{Spin-wave solution in homogeneous-DMI antiferromagnetic configuration}, the components of the spin-wave fluctuations can again be expressed in the form of Eqs. (\ref{components of spin-wave fluctuation for 2j sublattice}) and (\ref{components of spin-wave fluctuation for 2j+1 sublattice}). The degeneracy is lifted under homogeneous DMI. Eight spin-wave branches are obtained, four with positive frequencies and four with negative frequencies. Their amplitudes and initial phases, as well as the ellipticity angles [calculated from Eq. (\ref{ellipticity angle})] are listed in Tab. \ref{spin-wave parameters of homogeneous DMI-AFM configuration} for every branches of SWs. To better clarify these characteristics, we illustrate the precessional trajectories of the magnetization tip for all modes of the two sublattices in Tab. \ref{schematic of modes of homogeneous DMI-AFM configuration}. This schematic clearly presents their chirality, amplitude ratios and phase differences. Additionally, we plot the dispersions and the dependence of sublattice amplitude ratios on the wave number in Fig. \ref{dispersion and amplitude ratio of homogeneous DMI-AFM configuration} in the whole Brillouin zone.

In Tab. \ref{spin-wave parameters of homogeneous DMI-AFM configuration}, the eigenfrequencies of nutational and precessional SWs read
\begin{equation}
\omega_{n,p}^\pm = \frac{\sqrt{1 + \eta \Gamma \pm \sqrt{\left( 1 + \eta \Gamma \right)^2 - \eta^2 \Lambda_\pm^2}}}{\sqrt{2} \eta}, \label{dispersion of homogeneous DMI-AFM configuration}
\end{equation}
where $\Gamma = 4 \sqrt{\omega_E^2 + \omega_D^{c 2}}$, and $\Lambda_\pm = 4 [ ( \omega_E^2 + \omega_D^{c 2} ) - ( \omega_E^2 + \omega_D^2 ) \cos^2 ( k a \mp \varphi_d ) ]^{1/2}$, with $\omega_D^c$ defined in Eq. (\ref{critical DMI}), and $\varphi_d = \tan^{- 1} ( \omega_E, \omega_D )$. In Eq. (\ref{dispersion of homogeneous DMI-AFM configuration}), the upper and lower signs in front of the square root correspond to the $n$ and $p$ branches, respectively. Similar to Sec. \ref{secIVA}, $\Lambda_\pm$ is purely imaginary for some values of $k$ when $\omega_D > \omega_D^c$, yielding $\Lambda_\pm^2 < 0$. Consequently, $\omega_p^\pm$ becomes complex according to Eq. (\ref{dispersion of homogeneous DMI-AFM configuration}), indicating the instability of AFM configuration. Here, the degeneracy is lifted by the DMI. When $\omega_D = 0$, one has $\varphi_d = 0$, which further leads to $\omega_{n,p}^+ = \omega_{n,p}^-$, signifying the recovery of degeneracy for both nutational and precessional SWs.

In Tab. \ref{spin-wave parameters of homogeneous DMI-AFM configuration}, the parameters $\rho_{n,p}^\pm$ are given by
\begin{equation}
\rho_{n,p}^\pm = \pm \frac{2 \sqrt{\omega_E^2 + \omega_D^2} \cos \left( k a \mp \varphi_d \right)}{\eta \omega_{n,p}^{\pm 2} \pm \omega_{n,p}^\pm - 2 \sqrt{\omega_E^2 + \omega_D^{c 2}}}. \label{rho of homogeneous DMI-AFM configuration}
\end{equation}
Here, the leading $\pm$ sign and the $\pm$ sign in front of $\omega^{\pm}_{n,p}$ are associated with the subscripts $n$ and $p$, respectively, whereas the $\mp$ sign in the argument of the cosine is associated with the superscript $\pm$ of $\rho_{n,p}^\pm$. It can be verified that $\eta \omega_n^{\pm 2} + \omega_n^\pm - 2 \sqrt{\omega_E^2 + \omega_D^{c 2}} > 0$ and $- (\eta \omega_p^{\pm 2} - \omega_p^\pm - 2 \sqrt{\omega_E^2 + \omega_D^{c 2}}) > 0$. So, the signs of $\rho_{n,p}^\pm$ are determined by $\cos \left( k a \mp \varphi_d \right)$.

\vspace{-1em}

\subsubsection{Dispersion}

\begin{figure}[t]
\includegraphics[clip, trim = 0cm 2cm 0cm 2cm, width=0.45\textwidth]{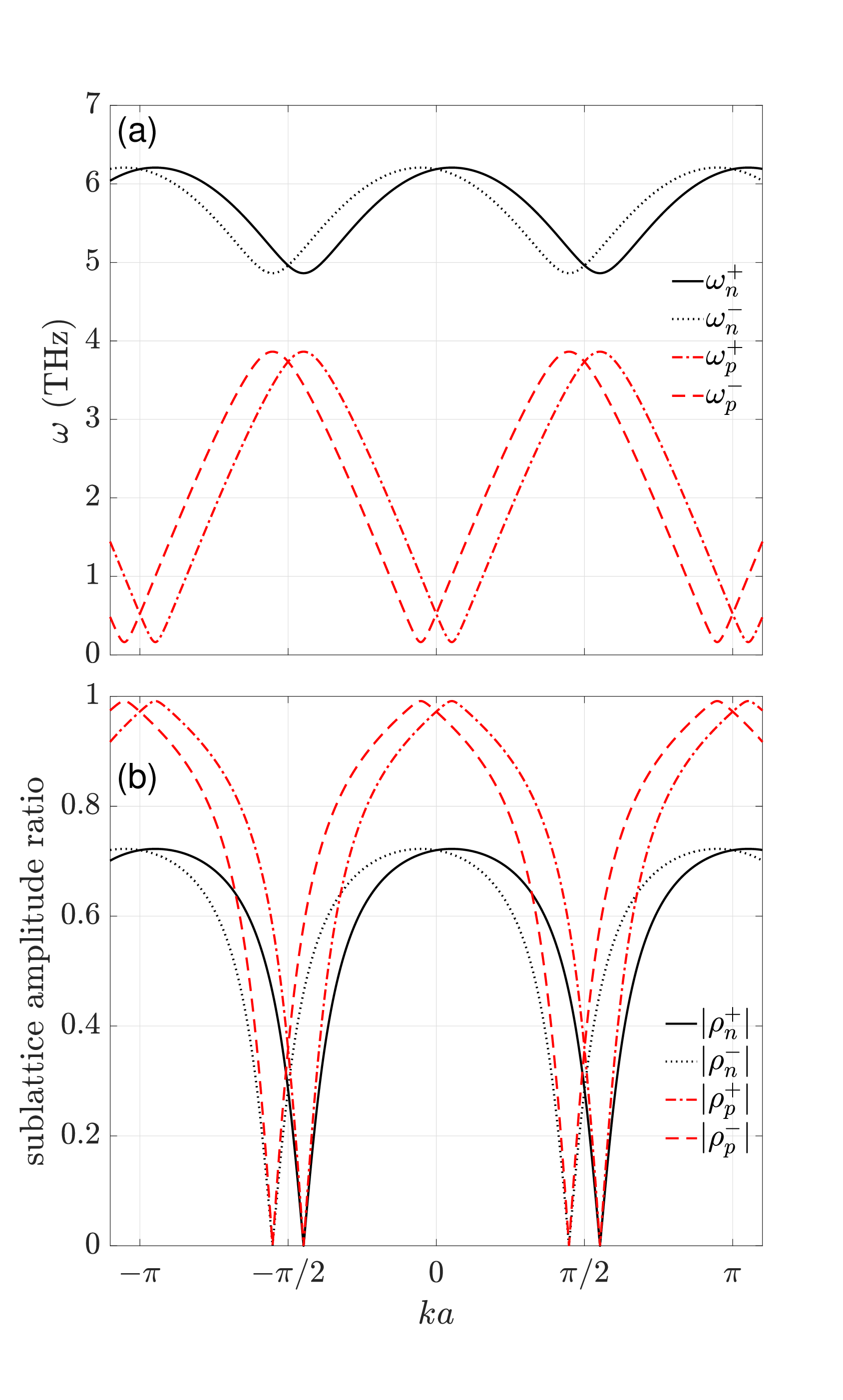}
\caption{(color online). Dispersion (a) and sublattice amplitude ratio (b) of the SWs on top of an AFM configuration for the case of homogeneous DMI. Here, all parameters remain identical to those in Fig. \ref{dispersion and amplitude ratio of staggered DMI-AFM configuration}.}
\label{dispersion and amplitude ratio of homogeneous DMI-AFM configuration}
\end{figure}

In Fig. \ref{dispersion and amplitude ratio of homogeneous DMI-AFM configuration}(a), the overall spin-wave spectrum exhibits no nonreciprocity. Nevertheless, individual spin-wave modes are nonreciprocal and emerge in pairs with opposite chirality. Their dispersion curves shift in opposite directions, while the combined spectrum remains reciprocal.

In contrast to Sec. \ref{secIVA}, which investigates SWs in the AFM configuration with staggered DMI, homogeneous DMI breaks the spin-wave degeneracy by shifting the dispersions along $k$ axis by $\pm \varphi_d/a$. This feature can be clearly observed by comparing Fig. \ref{dispersion and amplitude ratio of homogeneous DMI-AFM configuration}(a) and \ref{dispersion and amplitude ratio of staggered DMI-AFM configuration}(a). This distinction originates from their symmetry-breaking properties: homogeneous DMI breaks parity combined with time-reversal ($\mathcal{P} \mathcal{T}$) symmetry, whereas staggered DMI preserves it. Under spatial inversion, which is equivalent to interchanging the two magnetic sublattices, the homogeneous DMI term flips its sign and thereby breaks the $\mathcal{PT}$ symmetry of uniaxial antiferromagnets. By contrast, the staggered DMI term remains invariant under this operation.

Near the top of nutational dispersion and the bottom of precessional one ($k = \pm \varphi_d/a$), the frequencies expanded to second order in $k \pm \varphi_d/a$ read
\begin{equation}
\omega_{n,p}^\pm = \sqrt{\Delta_{n,p}^2 \pm v^2 \left( k \mp \varphi_d/a \right)^2},
\end{equation}
with the sign $\mp$ in front of $\varphi_d/a$ corresponding to the superscript $\pm$ of $\omega_{n,p}^\pm$, and the sign $\pm$ before $v^2$ corresponding to the subscript ($n,p$) of $\omega_{n,p}^\pm$. The frequency at the top of nutational dispersion $\Delta_n$ and the gap of the precessional dispersion $\Delta_p$ are given by Eq. (\ref{gap of staggered DMI-AFM configuration}). The characteristic velocity $v$ is given by Eq. (\ref{velocity of staggered DMI-AFM configuration}).

The nutational $\omega_n^+$ and $\omega_n^-$ branches reach their band minima at $k = (\pm \pi/2 + \varphi_d)/a$ and $k = (\pm \pi/2 - \varphi_d)/a$, respectively, while the corresponding precessional branches $\omega_p^+$ and $\omega_p^-$ reach their band maxima at the same wave numbers. The frequencies at the bottom of nutational band and the top of precessional band are still Eq. (\ref{bottom and top of bands of staggered DMI-AFM configuration}).

As $\eta \rightarrow 0$, $\omega^\pm_n$ diverges and vanishes. The dispersion displayed in Fig. \ref{dispersion and amplitude ratio of homogeneous DMI-AFM configuration}(a) is consistent with the result reported in Ref. [\onlinecite{Taku J Sato}].

\vspace{-1em}

\subsubsection{Sublattice amplitude ratio}

The amplitude ratio of SWs between two sublattices can be represented by $\vert \rho_{n,p}^\pm \vert$, which varies with $k$, as displayed in Fig. \ref{dispersion and amplitude ratio of homogeneous DMI-AFM configuration}(b). The sublattice amplitude ratios of the $\omega_n^+$- and $\omega_p^+$-branches reach their maxima at $k = \varphi_d/a$, whereas those of the $\omega_n^-$- and $\omega_p^-$-branches reach their maxima at $k = - \varphi_d/a$. In particular, the ratio is close to $1$ for the precessional mode. In the limit $\omega_K \ll \omega_E$ and $\omega_D \ll \omega_E$, $\rho_p^\pm$ reduces to the expression given in Eq. (\ref{rho for precessional mode at k = 0}). Owing to the relative smallness of $\omega_K$ and $\omega_D$, $\rho_p^0$ is about $1$ and the angle $\varphi_d$ is small. From this, we can conclude that the amplitudes of two sublattices are closest to each other near the long-wavelength limit. With increasing $k$ from $k = 0$, $\vert \rho_{n,p}^\pm \vert$ decreases gradually. The amplitude ratio $\rho_{n,p}^+$ vanishes at  $k = (\pm \pi/2 + \varphi_d)/a$, while $\rho_{n,p}^-$ vanishes at $k = (\pm \pi/2 - \varphi_d)/a$. This implies that only a single sublattice can be excited for short-wavelength SWs with $\lambda \approx 4 a$. In addition, it is worth mentioning that the amplitude ratios corresponding to positive- and negative-frequency branches are mutually reciprocal, as is evident by comparing the $A_\theta$ and $A_\phi$ data listed in Tab. \ref{spin-wave parameters of homogeneous DMI-AFM configuration}. This feature is also illustrated schematically in Tab. \ref{schematic of modes of homogeneous DMI-AFM configuration}.

\vspace{-1em}

\subsubsection{Chirality and phase}

Since we adopt local coordinate frames (shown in Tab. \ref{schematic of modes of homogeneous DMI-AFM configuration}) throughout spin-wave calculations, chirality is defined as the rotational sense about the static magnetization of each individual sublattice. Following this definition, SWs of two sublattices exhibit opposite signs of $\chi$ with $\vert \chi \vert = \pi/4$, indicating opposite chiralities that correspond to lefthanded and righthanded circular polarizations.

It is generally acknowledged that the nutational SWs are lefthanded and the precessional SWs are righthanded. Here, we find that one branch nutational SW exhibits lefthanded polarization, whereas the other branch is righthanded, and the same behavior applies to precessional SWs. Additionally, flipping the sign of eigenfrequency can reverse the chirality of SWs. As listed in Tab. \ref{spin-wave parameters of homogeneous DMI-AFM configuration}, the phase difference ($\delta_{\phi, l} - \delta_{\theta, l}$) of two components of $\delta \mathbf{m}_l$ remains unchanged under the sign reversal of $\omega$. Therefore, according to the definition of ellipticity angle in Eq. (\ref{ellipticity angle}), the sign of $\chi$ is the identical to that of $\omega$.

In Tab. \ref{schematic of modes of homogeneous DMI-AFM configuration}, we schematically illustrate the phases of all modes at a specific lattice site at a given instant in time. It reveals that the phase relation between the oscillations of two sublattices depends on the wave number. More specifically, for $0 < \vert k \vert a < \pi/2 - \varphi_d$ ($\lambda_\pm > 0$), the two sublattices oscillate in phase for nutation and in antiphase for precession; for $\pi/2 + \varphi_d < \vert k \vert a < \pi$ ($\lambda_\pm < 0$), the phase relations are reversed. In the narrow region $\pi/2 - \varphi_d < k a < \pi/2 + \varphi_d$ ($\lambda_+ > 0$ and $\lambda_- < 0$), the two sublattices oscillate in phase for the $\omega_n^+$ and $\omega_p^-$ modes and in antiphase for the $\omega_n^-$ and $\omega_p^+$ modes. The phase behavior is reversed in the interval $- \pi/2 - \varphi_d < k a < - \pi/2 + \varphi_d$ (where $\lambda_+ < 0$ and $\lambda_- > 0$).

\vspace{-1em}

\subsection{Spin waves on top of spiral configuration}

When $\omega_D > \omega_D^c$, the static configuration is spiral with a helical angle $\phi_s$ [Eq. (\ref{spiral angle})], as schematically displayed in Fig. \ref{magnetic configurations}(b). Applying the method presented in Sec. \ref{method}, the spin-wave modes obtained in Appendix \ref{Spin-wave solution in spiral configuration} can be expressed in the form of Eq. (\ref{components of spin-wave fluctuation}). The spin-wave spectrum consists of four branches: two positive-frequency branches and their two negative-frequency counterparts. Their amplitudes, initial phases, and ellipticity angles are listed in Tab. \ref{spin-wave parameters of spiral configuration} for all spin-wave branches.

The above results reveal that SWs based on the spiral configuration correspond exactly to the low-frequency branch of nutational and precessional SWs in the canted magnetic configuration. Accordingly, their dispersion relations, oscillation amplitudes, phase characteristics and ellipticity angles are identical to those of the $\omega_{n,p}^-$ spin-wave branches within the canted configuration, which is verified by a comparison between Table \ref{spin-wave parameters of spiral configuration} and the data of $\omega_{n,p}^-$ branches listed in Tab. \ref{spin-wave parameters of canted configuration}.

This correspondence is not a coincidence: it originates from the invariance of dispersion relations under local unitary transformations \cite{J. J. Sakurai}. Transforming the magnetic energy with a homogeneous DMI [Eq. (\ref{reduced magnetic energy})] into the local frame defined by Eq. (\ref{local frame}) for the spiral state yields
\begin{eqnarray}
E_{spiral} &=& \sum_l \Big[ \sqrt{\omega_E^2 + \omega_D^2} \left( m_l^r m_{l + 1}^r + m^\phi_l m^\phi_{r + 1} \right) \notag \\ && + \omega_E  m_l^\theta m_{l + 1}^\theta - \omega_K \left( m_l^\theta \right)^2 \Big], \label{transformed energy of spiral configuration}
\end{eqnarray}
which corresponds to a FM spin chain with anisotropic exchange and uniaxial magnetic anisotropy, and gives the spin-wave dispersions of spiral configuration. In the local frame defined by Eq. (\ref{local frame}) and the canted state, the magnetic energy with a staggered DMI [Eq. (\ref{reduced magnetic energy})] differs from $E_{spiral}$ only by an extra minus sign in the first term. By further rotating $\mathbf{m}_l$ by $\pi$ around the local $\mathbf{e}^l_\theta$ axis, we attain $E_{canted} = E_{spiral}$, which ensures that the spin-wave dispersion of the canted configuration (its $\omega_n^-$- and $\omega_p^-$-branches) is identical to that of the spiral configuration.

To understand the emergence of $\omega_{n,p}^+$ spin-wave branch on the canted configuration, one can invoke band folding \cite{Roald Hoffmann}. The equivalent ferromagnetic chain described by Eq. (\ref{transformed energy of spiral configuration}) supports the $\omega_{n,p}^-$ dispersion defined in the extended Brillouin zone $k \in [- \pi/a, \pi/a]$. In contrast, the original canted configuration has two magnetic atoms per primitive cell, so its first Brillouin zone is reduced to $k \in [- \pi/(2 a), \pi/(2 a)]$. By folding the $\omega_{n,p}^-$ dispersion from the outer intervals $[- \pi/a, - \pi/(2 a)]$ and $[\pi/(2 a), \pi/a]$ into this reduced zone, one obtain the $\omega_{n,p}^+$ branch. Specifically, the dispersions in $[- \pi/a, - \pi/(2 a)]$ is translated by $\pi/a$, and that in $[\pi/(2 a), \pi/a]$ by $- \pi/a$. This yields $\omega_{n,p}^+ (k) = \omega_{n,p}^- (k \pm \pi/a)$, a relation that can also be verified using Eq. (\ref{omega of canted configuration}).

\begin{table}[t]
\caption{Amplitudes, phases, and ellipticity angles of the SWs on top of a spiral configuration for homogeneous DMI. Here, $X_{n,p} = \frac{1}{2} \arcsin \left( \frac{2 \rho_{n,p} \omega_{n,p}}{1 + \rho_{n,p}^2 \omega_{n,p}^2} \right)$.}
\begin{tblr}{
  colspec = {*{13}{c}},
  rows = {valign = m}
}
\hline \hline
& & $\omega_n$ & $- \omega_n$ & & $\omega_p$ & $- \omega_p$ \\
\hline
$\rho_{\theta, l}$ & & $\rho_n$ & $\rho_n$ & & $\rho_p$ & $\rho_p$ \\
$\rho_{\phi, l}$ & & $\frac{1}{\omega_n}$ & $\frac{1}{\omega_n}$ & & $\frac{1}{\omega_p}$ & $\frac{1}{\omega_p}$ \\
$\delta_{\theta, l}$ & & $\pi$ & $\pi$ & & $0$ & $0$ \\
$\delta_{\phi, l}$ & & $\frac{\pi}{2}$ & $- \frac{\pi}{2}$ & & $\frac{\pi}{2}$ & $- \frac{\pi}{2}$ \\
$\chi_l$ & & $- X_n$ & $- X_n$ & & $X_p$ & $X_p$ \\
 \hline \hline
\end{tblr}
\label{spin-wave parameters of spiral configuration}
\end{table}

\vspace{-1em}

\section{Discussions} \label{Discussions}

We first compare the spin-wave degeneracies in the AFM configuration under homogeneous and staggered DMIs. The degeneracy is lifted in the presence of homogeneous DMI but preserved under staggered DMI, which originates from the intrinsic symmetry difference between the two DMI forms. It is well established that $\mathcal{PT}$ symmetry guarantees the double degeneracy of SWs in uniaxial AFM systems. For homogeneous DMI, interchanging the two magnetic sublattices flips the sign of the DMI term, which breaks $\mathcal{P}$ symmetry, and further destroys the overall $\mathcal{PT}$-symmetry. Consequently, the two originally degenerate spin-wave dispersions shift oppositely along the $k$-axis, as illustrated in Fig. \ref{dispersion and amplitude ratio of homogeneous DMI-AFM configuration}(a). By contrast, staggered DMI retains intact $\mathcal{PT}$ symmetry, such that the spin-wave degeneracy remains unaltered, as depicted in Fig. \ref{dispersion and amplitude ratio of staggered DMI-AFM configuration}(a).

Following the same symmetry-based reasoning, we compare the dispersions of SWs atop the collinear AFM and canted configurations, as shown in Figs. \ref{dispersion and amplitude ratio of staggered DMI-AFM configuration}(a) and \ref{dispersion and ellipticity angle of canted configuration}(a). In the AFM configuration, the system preserves the $\mathcal{PT}$ symmetry. Combining with the $\mathcal{PT}$ symmetric staggered DMI, the spin-wave degeneracy is not lifted. In contrast, the canted configuration breaks both $\mathcal{P}$ and the $\mathcal{T}$ symmetries, even though the staggered DMI itself remains $\mathcal{PT}$ symmetric. As a result, the originally twofold degenerate spin-wave spectrum is completely lifted: near the zone center ($k = 0$) the two branches split along the $\omega$-axis, much like the separation into an acoustic (gapless) and an optical (gapped) branch; near the zone boundary [$k = \pm \pi/(2 a)$] they shift along the k-axis. The former splitting arises directly from the $\mathcal{T}$ symmetry breaking. The latter shifting originates from the $\mathcal{P}$ symmetry breaking. In the presence of $\mathcal{P}$ symmetry, the dispersion must satisfy $\omega (k_b + \delta k) = \omega (k_b - \delta k)$ with $k_b = \pi/(2 a)$, which forces the group velocity at the zone boundary to vanish and pins the extrema at this high-symmetry point. Once the $\mathcal{P}$ symmetry is broken, the group velocities at the zone boundary become nonzero and opposite in sign for the two branches; consequently, their extrema are no longer pinned at the high-symmetry point but shift in opposite directions, leading to a crossing and an apparent lateral displacement, as shown in Fig. \ref{dispersion and ellipticity angle of canted configuration zoomin}.

After the degeneracy is lifted, each dispersion branch remains an even function of $k$. This can be understood from a symmetry perspective: although $\mathcal{T}$ and $\mathcal{P}$ are individually broken, the canted state still preserves the anti-unitary symmetry that combines time reversal with a $\pi$ rotation about the $z$ axis, $\mathcal{T}$$R_z(\pi)$. Under this operation, each magnon branch is mapped onto itself at the opposite $k$, which enforces $\omega(k) = \omega(- k)$ for every branch.

Second, there exists a prevalent yet misleading notion that reversing the sign of frequency will invert the chirality of SWs. This conclusion holds true only for several simple spin-wave forms, such as the spin fluctuation $\delta \mathbf{m} = \delta m_x \cos \omega t \mathbf{e}_x + \delta m_y \sin \omega t \mathbf{e}_y$. Upon replacing $\omega$ with $- \omega$, the chirality of $\delta \mathbf{m}$ switches from right-handed to left-handed. However, the general solutions, as exemplified in Eq. (\ref{components of spin-wave fluctuation}), are not that straightforward. The initial phases of each component in the spin-wave solutions, and consequently the phase differences between these components, generally differ between positive-frequency and negative-frequency solutions, as illustrated in Tabs. \ref{spin-wave parameters of staggered DMI-AFM configuration}, \ref{spin-wave parameters of canted configuration}, \ref{spin-wave parameters of homogeneous DMI-AFM configuration}, and \ref{spin-wave parameters of spiral configuration}. Therefore, it is essential to adhere to the general principle that the chirality of the superposition of two perpendicular harmonic oscillations depends on their phase difference. The ellipticity angle $\chi$ defined in Eq. (\ref{ellipticity angle}) takes into account both the phase different between spin-wave components and the sign of frequency. The sign of $\chi$ characterizes the spin-wave chirality, while its magnitude denotes the polarization ellipticity. This definition covers all polarization states including linear, elliptical and circular polarizations, as well as all possible orientations of the trajectory traced by the magnetization vector tip. In the discussed cases, flipping the frequency sign exerts no influence on the chirality of SWs in spiral, canted and staggered-DMI AFM configurations. By contrast, such an operation reverses the chirality of SWs in homogeneous-DMI AFM configuration.

We note that, building upon the lengths of the principal axes of the elliptical precession trajectory, Ref. [\onlinecite{Yutian Wang}] recently proposed a signed ellipticity parameter to quantify the polarization and chirality of SWs. However, this definition only captures axis-aligned elliptical polarization states (with principal axes aligned with the coordinate axes) and linear polarization along the $x$-axis, failing to describe elliptical and linear polarizations with arbitrary orientations---i.e., all possible polarization states.

Third, we adopt the classical linearization approach to derive spin-wave solutions. By solving the eigenvalues and eigenvectors of the linearized dynamic equations, explicit analytical expressions for SWs can be directly acquired from the corresponding eigenvectors. In this way, the key spin-wave characteristics, including dispersion relations, amplitudes, phases, chiralities, and polarization states, can be readily extracted from the eigenvalues and eigenvectors. The proposed method therefore provides a general and efficient framework for characterizing inertial spin waves. The quantum formalism, based on the Holstein-Primakoff transformation, Fourier transformation, and Bogoliubov transformation, offers a powerful and systematic description of magnonic excitations. However, when the goal is to obtain explicit spin-wave amplitudes, phases, chiralities, and polarization states, these quantities may be less transparent in the Bogoliubov representation and require additional reconstruction. In this respect, the classical linearization approach provides a more direct route to chirality and polarization, while remaining fully complementary to the quantum approach. The latter is particularly advantageous for addressing intrinsically quantum phenomena, such as magnon-magnon interactions, nonlinear effects, magnon Bose-Einstein condensation, and related many-body processes.

Fourth, most nutation SWs [as shown in Figs. \ref{dispersion and amplitude ratio of staggered DMI-AFM configuration}(a), \ref{dispersion and ellipticity angle of canted configuration}(a) and \ref{dispersion and amplitude ratio of homogeneous DMI-AFM configuration}(a)] exhibit backward propagation characteristics (i.e., group velocity antiparallel to the wavevector) in the long-wavelength regime. This differs from classical backward volume magnetostatic SWs \cite{R. W. Damon}, which originate from magnetostatic dipole interactions. A similar backward feature was also reported for dipolar-governed nutation SWs in thin films \cite{Mikhail Cherkasskii PRB103}. Such conventional SWs show backward behavior only along specific propagation directions (e.g., near the in-plane magnetic field or in coupled layer structures) and within the gigahertz frequency range. For nutation SWs governed by the inertial LLG equation, however, the backward characteristic is nearly universal: the group velocity is oriented opposite to the wavevector for almost all propagation directions. Inertial nutation SWs therefore constitute a distinct category of backward SWs. They support backward energy transport without specially designed magnetostatic boundary conditions and open new opportunities for developing ultrafast terahertz spin-wave devices.

Finally, nonreciprocal effects in chiral magnets have previously been reported to emerge only at small wave numbers near the Brillouin zone center. By contrast, Figs. \ref{dispersion and ellipticity angle of canted configuration}(a),  \ref{dispersion and ellipticity angle of canted configuration}(c), \ref{dispersion and ellipticity angle of canted configuration zoomin} and \ref{dispersion and amplitude ratio of homogeneous DMI-AFM configuration}(a) reveal nonreciprocity extending to the Brillouin zone boundary.

\vspace{-1em}

\section{Conclusions} \label{Conclusions}

In this work, we propose a straightforward scheme for calculating inertial SWs in both uniform and nonuniform magnetic structures. After linearizing the inertial Landau-Lifshitz equation and recasting it into a set of first-order differential equations, we adopt standard linear-algebra techniques to directly solve for all eigenvalues and eigenvectors of inertial SWs. The corresponding amplitudes, dispersion relations, phases, chirality and polarizations can then be readily obtained. We apply this method to uniaxial AFMs with staggered and homogeneous DMIs, where uniform AFM, canted and spiral configurations emerge depending on the magnitude of DMI. Around these equilibrium magnetic configurations, we derive compact formulae for the dispersion relation, the sublattice amplitude ratio and the ellipticity angle (which jointly define chirality and polarization).

Detailed analysis of the calculated results uncovers key characteristics of inertial SWs in several magnetic configurations, as follows: (1) weak staggered DMI fails to lift the degeneracy of both nutational and precessional SWs, whereas weak homogeneous DMI does so. Under strong DMIs, the magnetic configurations are nonuniform and the SWs are fully non-degenerate. (2) In the long-wavelength regime, nutational SWs are generally backward waves with negative group velocities. For canted and spiral configurations, the group velocity of one nutational branch flips sign at a certain inertial relaxation time ($\eta$), producing flat bands. Our calculations show that under identical DMI strength, nutational and precessional SWs in spiral configuration coincide with the low-frequency branches of their counterparts in canted configuration. Band folding and dispersion-preserving unitary transformations naturally account for this equivalence. (3) The sublattice amplitude ratio of uniform AFM configurations strongly depends on the wave number and spin-wave branch, while SWs in canted configuration possess identical sublattice amplitude and polarization. (4) In the uniform AFM configurations, circularly polarized SWs with opposite chiralities always emerge in pairs, regardless of whether they are nutational or precessional SWs. For canted and spiral configurations, nutational SWs are always left-handed, while precessional SWs are always right-handed. (5) Spin wave polarization is insensitive to wave number for uniform AFM configurations, but strongly dependent on wave number for canted and spiral configurations.

The proposed methodology can be generalized to investigate inertial SWs in a broad range of magnetic configurations. Moreover, our findings provide a comprehensive characterization of inertial SWs in uniaxial DMI AFMs, which not only enriches the fundamental understanding of magnetic inertial effects but also offers valuable insights for the applied research of ultrafast magnonics.

\vspace{-1em}

\section{Acknowledgments}

This work was supported by the NSF of Changsha City (Grant No. kq2208008), the NSF of Hunan Province (Grant No. 2023JJ30116), the key Program of Education Bureau of Hunan Province (Grant No.24A0494) and the Regional Joint Funds of the NSF of Hunan Province (Grant No.2024JJ7312).

\appendix

\vspace{-1em}

\section{Expansion of effective field} \label{expansion of effective field}

In this section, we give the linear expansion of Eq. (\ref{effective field}). Inserting the spin-wave ansatz expressed by Eqs. (\ref{spin-wave ansatz}) and (\ref{spin-wave fluctuation}), the magnetic energy can be written as a quadratic function of $m_{\theta, l}$ and $m_{\phi, l}$. The zeroth-order contribution to the effective field is then given by
\begin{equation}
\mathbf{h}^0_l = - \left( \frac{\partial E}{\partial m_{\theta, l}} \right)_0 \mathbf{e}_\theta^l - \left( \frac{\partial E}{\partial m_{\phi, l}} \right)_0 \mathbf{e}_\phi^l, \label{zero-order effective field}
\end{equation}
where the subscript `$0$' denotes evaluation at the equilibrium state. Only considering the nearest-neighbor exchange coupling, the linear-order term of the effective field reads
\begin{widetext}
\vspace{-1em}
\begin{eqnarray}
\delta \mathbf{h}_l &=& - \left \{ \left[ \left( \frac{\partial^2 E}{\partial m_{\theta, l}^2} \right)_0 m_{\theta, l} + \left( \frac{\partial^2 E}{\partial m_{\theta, l} \partial m_{\phi, l}} \right)_0 m_{\phi, l} \right] + \left[ \left( \frac{\partial^2 E}{\partial m_{\theta, l} \partial m_{\theta,l + 1}} \right)_0 m_{\theta, l + 1} + \left( \frac{\partial^2 E}{\partial m_{\theta, l} \partial m_{\phi, l + 1}} \right)_0 m_{\phi, l + 1} \right]
\right. \notag \\
&& + \left. \left[ \left( \frac{\partial^2 E}{\partial m_{\theta, l} \partial m_{\theta,l - 1}} \right)_0 m_{\theta, l - 1} + \left( \frac{\partial^2 E}{\partial m_{\theta, l} \partial m_{\phi, l - 1}} \right)_0 m_{\phi, l - 1} \right] \right \} \mathbf{e}_\theta^l \notag \\
&& - \left \{ \left[ \left( \frac{\partial^2 E}{\partial m_{\phi, l}^2} \right)_0 m_{\phi, l} + \left( \frac{\partial^2 E}{\partial m_{\phi, l} \partial m_{\theta, l}} \right)_0 m_{\theta, l} \right] + \left[ \left( \frac{\partial^2 E}{\partial m_{\phi, l} \partial m_{\phi, l + 1}} \right)_0 m_{\phi, l + 1} + \left( \frac{\partial^2 E}{\partial m_{\phi, l} \partial m_{\theta, l + 1}} \right)_0 m_{\theta, l + 1} \right] \right. \notag \\
&& + \left. \left[ \left( \frac{\partial^2 E}{\partial m_{\phi, l} \partial m_{\phi, l - 1}} \right)_0 m_{\phi, l + 1} + \left( \frac{\partial^2 E}{\partial m_{\phi, l} \partial m_{\theta, l - 1}} \right)_0 m_{\theta, l - 1} \right] \right \} \mathbf{e}_\phi^l. \label{linear-order effective field}
\end{eqnarray}
\end{widetext}
$\delta \mathbf{h}_l$ can be calculated by the derivatives of magnetic energy with respect to the components of SW fluctuation. Eqs. (\ref{zero-order effective field}) and (\ref{linear-order effective field}) yield the expansion given in Eq. (\ref{linearized effective field}).

\vspace{-1em}

\section{SW solution in staggered-DMI antiferromagnetic configuration} \label{Spin-wave solution in staggered-DMI antiferromagnetic configuration}

When $\omega_D < \omega_D^c$, the AFM configuration is preferred, which is double degenerate. Without loss of generality, we chose $\mathbf{m}^0_{2 j} = - \mathbf{e}_z$ and $\mathbf{m}^0_{2 j + 1} = \mathbf{e}_z$. In the coordinate system shown in Fig. \ref{magnetic configurations}(a), the parametrization given in Eq. (\ref{local frame}) becomes singular when  $\mathbf{m}^0_l$ is aligned with the $z$ axis, since $\phi_l^0$ is ill-defined.  To resolve this singularity, we instead employ a rotated frame in which $\mathbf{m}^0_{2 j} = - \mathbf{e}_x$ and $\mathbf{m}^0_{2 j + 1} = \mathbf{e}_x$, corresponding to $\theta^0_{2 j} = \theta^0_{2 j + 1} = \pi/2$, $\phi_{2 j}^0 = \pi$ and $\phi_{2 j + 1}^0 = 0$. Then we combine this equilibrium magnetization with the spin-wave fluctuations in Eq. (\ref{spin-wave fluctuation}) to obtain the total magnetization. Substituting the resultant expression into the reduced magnetic energy in Eq. (\ref{reduced magnetic energy}) for staggered DMI (where $\mathbf{e}_z$ is replaced by $\mathbf{e}_x$), along with the coordinate transformation given in Eq. (\ref{local frame}), one can explicitly calculate all matrices in Eq. (\ref{Hessian matrix}) as follows: $\mathcal{H}_l = 2 \operatorname{diag} [\omega_E + \omega_K, \omega_E + \omega_K]$, and
\begin{equation}
\mathcal{H}_{l \pm 1} = \! \left( \! \begin{array}{cc} \omega_E & - \! \left( \! - 1 \right)^l \omega_D \cos \phi_l^0 \\ - \! \left( \! - 1 \right)^l \omega_D \cos \phi_l^0 & - \omega_E \end{array} \! \right) \!. \label{Hessian matrix of staggered DMI-AFM configuration}
\end{equation}

Next, the parameter matrixes $\mathcal{M}_l$ and $\mathcal{M}_{l \pm 1}$ in Eq. (\ref{spin-wave eigen-equation}) can be assembled using Eqs. (\ref{symplectic unit matrix}), (\ref{parameter matrix at site l}) and (\ref{parameter matrix at site l +- 1}). Since $\mathcal{H}_{l \pm 1}$ depends on $\phi_l^0$ ($\phi_{2 j}^0 = \pi$ and $\phi_{2 j + 1}^0 = 0$), it follows that $\mathcal{H}_{l \pm 1}$ is identical across all lattice sites within each sublattice, while distinct between two different sublattices. Thus, the complex amplitudes in the plane-wave ansatz [Eq. (\ref{plane-wave solution})] should be extended as $(\mathcal{A}^e_\theta, \mathcal{A}^e_\phi, \mathcal{A}^o_\theta, \mathcal{A}^o_\phi)^T$, with $\mathcal{A}^{e(o)}_\theta$ and $\mathcal{A}^{e(o)}_\phi$ being the uniform amplitudes of even (odd) sites, and $T$ denoting the transpose of matrix. Then the spin-wave equations [Eq. (\ref{spin-wave eigen-equation})] for two sublattices can be written as
\begin{equation}
\mathcal{M}_{2 j} \left( \! \begin{array}{c} \mathcal{A}^e_\theta \\[0.5em] \mathcal{A}^e_\phi \\[0.5em] \omega \mathcal{A}^e_\theta \\[0.5em] \omega \mathcal{A}^e_\phi \end{array} \! \right) + \widetilde{\mathcal{M}}_{2 j} \! \left( \! \begin{array}{c} \mathcal{A}^o_\theta \\[0.5em] \mathcal{A}^o_\phi \\[0.5em] \omega \mathcal{A}^o_\theta \\[0.5em] \omega \mathcal{A}^o_\phi \end{array} \! \right) \! = \omega \left( \! \begin{array}{c} \mathcal{A}^e_\theta \\[0.5em] \mathcal{A}^e_\phi \\[0.5em] \omega \mathcal{A}^e_\theta \\[0.5em] \omega \mathcal{A}^e_\phi \end{array} \! \right), \label{matrix equation 1}
\end{equation}
and
\begin{equation}
\mathcal{M}_{2 j + 1} \left( \! \begin{array}{c} \mathcal{A}^o_\theta \\[0.5em] \mathcal{A}^o_\phi \\[0.5em] \omega \mathcal{A}^o_\theta \\[0.5em] \omega \mathcal{A}^o_\phi \end{array} \! \right) + \widetilde{\mathcal{M}}_{2 j + 1} \! \left( \! \begin{array}{c} \mathcal{A}^e_\theta \\[0.5em] \mathcal{A}^e_\phi \\[0.5em] \omega \mathcal{A}^e_\theta \\[0.5em] \omega \mathcal{A}^e_\phi \end{array} \! \right) \! = \omega \left( \! \begin{array}{c} \mathcal{A}^o_\theta \\[0.5em] \mathcal{A}^o_\phi \\[0.5em] \omega \mathcal{A}^o_\theta \\[0.5em] \omega \mathcal{A}^o_\phi \end{array} \! \right), \label{matrix equation 2}
\end{equation}
where $\widetilde{\mathcal{M}}_l = e^{i k a} \mathcal{M}_{l + 1} + e^{- i k a} \mathcal{M}_{l - 1}$. Combining these two matrixes equations (\ref{matrix equation 1}) and (\ref{matrix equation 2}), and calculating the parameter matrixes $\mathcal{M}_l$ and $\mathcal{M}_{l \pm 1}$, we derive the matrix-form eigen equation for SWs as follows:
\enlargethispage{2\baselineskip}
\begin{equation}
\mathcal{M} \left( \begin{array}{c} \mathcal{A}^e_\theta \\[0.5em] \mathcal{A}^e_\phi \\[0.5em] \mathcal{A}^o_\theta \\[0.5em] \mathcal{A}^o_\phi \\[0.5em]\omega \mathcal{A}^e_\theta \\[0.5em] \omega \mathcal{A}^e_\phi \\[0.5em] \omega \mathcal{A}^o_\theta \\[0.5em] \omega \mathcal{A}^o_\phi \end{array} \right)
= \omega \left( \begin{array}{c} \mathcal{A}^e_\theta \\[0.5em] \mathcal{A}^e_\phi \\[0.5em] \mathcal{A}^o_\theta \\[0.5em] \mathcal{A}^o_\phi \\[0.5em]\omega \mathcal{A}^e_\theta \\[0.5em] \omega \mathcal{A}^e_\phi \\[0.5em] \omega \mathcal{A}^o_\theta \\[0.5em] \omega \mathcal{A}^o_\phi \end{array} \right), \label{spin-wave eigen-equation for bipartite configuration}
\end{equation}
\par\vfill\eject\noindent
where the parameter matrix
\begin{equation}
\mathcal{M} = \left( \begin{array}{cc} O_4 & I_4 \\[1em] \mathcal{M}_{21} & \mathcal{M}_{22} \end{array} \right), \label{combined parameter matrix for bipartite configuration}
\end{equation}
with $O_4$ and $I_4$ being the $4 \times 4$ zero matrix and identity matrix, respectively. In Eq. (\ref{combined parameter matrix for bipartite configuration}),
\begin{equation}
\mathcal{M}_{22} = \frac{i}{\eta} \left( \begin{array}{cccc}
0 & 1 & 0 & 0 \\
- 1 & 0 & 0 & 0 \\
0 & 0 & 0 & - 1 \\
0 & 0 & 1 & 0
\end{array} \right),
\end{equation}
and
\begin{equation}
\mathcal{M}_{21} = \frac{2}{\eta} \left( \! \begin{array}{cccc}
\omega_E \! + \! \omega_K & 0 & \omega_E cs & \omega_D cs \\
0 & \omega_E \! + \! \omega_K & \omega_D cs & - \omega_E cs \\
\omega_E cs & \omega_D cs & \omega_E \! + \! \omega_K & 0 \\
\omega_D cs & - \omega_E cs & 0 & \omega_E \! + \! \omega_K
\end{array} \! \right), \label{M12 of staggered DMI-AFM configuration}
\end{equation}
where we have introduced $\cos k a = cs$ to shorten the expressions.

After solving Eq. (\ref{spin-wave eigen-equation for bipartite configuration}), the eigenfrequencies and eigenvectors can be arranged in matrix form as
\begin{equation}
\mathcal{M} \mathcal{V} = \Omega \mathcal{V}, \label{eigenvalue matrix equation}
\end{equation}
where $\Omega$ is the diagonal matrix of the eigenvalues, i.e.
\begin{equation}
\Omega = \operatorname{diag} \left[ \omega_n, \omega_n, - \omega_n, - \omega_n, \omega_p, \omega_p, - \omega_p, - \omega_p \right].
\vspace{6pt}
\end{equation}
Here, $\omega_n$ and $\omega_p$ [unified in Eq. (\ref{omega of staggered DMI-AFM configuration})] are the nutational and precessional frequencies, respectively. The eigenvector matrix $\mathcal{V}$ in Eq. (\ref{eigenvalue matrix equation}) is written as
\begin{widetext}
\begin{equation}
\mathcal{V} = \left( \begin{array}{cccccccc}
\frac{\rho_n}{\omega_n} e^{- i \varphi_0} & - \frac{e^{i \varphi_0}}{\rho_n \omega_n} & \frac{\rho_n}{\omega_n} e^{i \varphi_0} & - \frac{e^{- i \varphi_0}}{\rho_n \omega_n} & \frac{\rho_p}{\omega_p} e^{i \varphi_0} & - \frac{e^{- i \varphi_0}}{\rho_p \omega_p} & \frac{\rho_p}{\omega_p} e^{- i \varphi_0} & - \frac{e^{i \varphi_0}}{\rho_p \omega_p} \\[2.5mm]
i \frac{\rho_n}{\omega_n} e^{- i \varphi_0} & i \frac{e^{i \varphi_0}}{\rho_n \omega_n} & - i \frac{\rho_n}{\omega_n} e^{i \varphi_0} & - i \frac{e^{- i \varphi_0}}{\rho_n \omega_n} & - i \frac{\rho_p}{\omega_p} e^{i \varphi_0} & - i \frac{e^{- i \varphi_0}}{\rho_p \omega_p} & i \frac{\rho_p}{\omega_p} e^{- i \varphi_0} & i \frac{e^{i \varphi_0}}{\rho_p \omega_p} \\[2.5mm]
\frac{1}{\omega_n} & - \frac{1}{\omega_n} & \frac{1}{\omega_n} & - \frac{1}{\omega_n} & - \frac{1}{\omega_p} & \frac{1}{\omega_p} & - \frac{1}{\omega_p} & \frac{1}{\omega_p} \\[2.5mm]
- \frac{i}{\omega_n} & - \frac{i}{\omega_n} & \frac{i}{\omega_n} & \frac{i}{\omega_n} & - \frac{i}{\omega_p} & - \frac{i}{\omega_p} & \frac{i}{\omega_p} & \frac{i}{\omega_p} \\[2.5mm]
\rho_n e^{- i \varphi_0} & - \frac{e^{i \varphi_0}}{\rho_n} & - \rho_n e^{i \varphi_0} & \frac{e^{- i \varphi_0}}{\rho_n} & \rho_p e^{i \varphi_0} & - \frac{e^{- i \varphi_0}}{\rho_p} & - \rho_p e^{- i \varphi_0} & \frac{e^{i \varphi_0}}{\rho_p} \\[2.5mm]
i \rho_n e^{- i \varphi_0} & i \frac{e^{i \varphi_0}}{\rho_n} & i \rho_n e^{i \varphi_0} & i \frac{e^{- i \varphi_0}}{\rho_n} & - i \rho_p e^{i \varphi_0} & - i \frac{e^{- i \varphi_0}}{\rho_p} & - i \rho_p e^{- i \varphi_0} & - i \frac{e^{i \varphi_0}}{\rho_p} \\[2.5mm]
1 & - 1 & - 1 & 1 & - 1 & 1 & 1 & - 1 \\[2.5mm]
- i & - i & - i & - i & - i & - i & - i & - i
\end{array} \right),
\end{equation}
\end{widetext}
with $\varphi_0 = \tan^{- 1} (\omega_E, \omega_D)$. Here, $\rho_n$ and $\rho_p$ are expressed collectively as Eq. (\ref{rho of staggered DMI-AFM configuration}). The columns of $\mathcal{V}$ correspond sequentially to the eigenvectors associated with the frequencies given in the diagonal matrix of the eigenfrequencies $\Omega$ for circularly polarized SWs. From $\mathcal{V}$, we then extract the complex amplitudes of SWs, i.e. $\mathcal{A}^e_\theta$, $\mathcal{A}^e_\phi$, $\mathcal{A}^o_\theta$, and $\mathcal{A}^o_\phi$. Substituting these amplitudes into the plane-wave solutions in Eq. (\ref{plane-wave solution}) and taking the real parts yields the explicit expressions for spin-wave excitations, as formulated in Eqs. (\ref{components of spin-wave fluctuation for 2j sublattice}) and (\ref{components of spin-wave fluctuation for 2j+1 sublattice}). All relevant parameters are summarized in Tab. \ref{spin-wave parameters of staggered DMI-AFM configuration}.

\vspace{-1em}

\section{SW solution in canted configuration} \label{Spin-wave solution in canted configuration}

When $\omega_D > \omega_D^c$, the canted configuration [sketched in Fig. \ref{magnetic configurations}(c)] is energetically favorable at equilibrium and doubly degenerate. We chose $\mathbf{m}^0_{2 j} = - \cos \phi_c \mathbf{e}_x + \sin \phi_c \mathbf{e}_y$ and $\mathbf{m}^0_{2 j + 1} = \cos \phi_c \mathbf{e}_x + \sin \phi_c \mathbf{e}_y$, i.e. $\theta^0_{2 j} = \theta^0_{2 j + 1} = \pi/2$, $\phi^0_{2 j} = \pi - \phi_c$, and $\phi^0_{2 j + 1} = \phi_c$, with $\phi_c$ being Eq. (\ref{canted angle}). Then we combine this equilibrium magnetization with the spin-wave fluctuations in Eq. (\ref{spin-wave fluctuation}) to obtain the total magnetization. Substituting the resultant expression into the reduced magnetic energy in Eq. (\ref{reduced magnetic energy}) for staggered DMI, along with the coordinate transformation given in Eq. (\ref{local frame}), one can explicitly calculate all matrices in Eq. (\ref{Hessian matrix}) as follows: $\mathcal{H}_l = 2 \operatorname{diag} [ \omega_E \cos \phi^0_l - ( - 1 )^l \omega_D \sin \phi^0_l - \omega_K, \omega_E \cos \phi^0_l - ( - 1 )^l \omega_D \sin \phi^0_l ]$, and $\mathcal{H}_{l \pm 1} = \operatorname{diag} [ \omega_E, - \omega_E \cos \phi^0_l + ( - 1 )^l \omega_D \sin \phi^0_l ]$.

Given that the canted configuration possesses the identical magnetic period as the AFM phase, following the routine in Sec. \ref{Spin-wave solution in staggered-DMI antiferromagnetic configuration} we obtain the similar spin-wave eigen equation described by Eqs. (\ref{spin-wave eigen-equation for bipartite configuration}) and (\ref{combined parameter matrix for bipartite configuration}) with
\vspace{-1em}
\begin{equation}
\mathcal{M}_{21} = \frac{2}{\eta} \left( \! \begin{array}{cccc}
\widetilde{\omega}_E \! - \! \omega_K & 0 & \omega_E cs & 0 \\
0 & \widetilde{\omega}_E & 0 & - \widetilde{\omega}_E cs \\
\omega_E cs & 0 & \widetilde{\omega}_E \! - \! \omega_K & 0 \\
0 & - \widetilde{\omega}_E cs & 0 & \widetilde{\omega}_E
\end{array} \! \right), \label{M matrix for canted configuration}
\end{equation}
where $\widetilde{\omega}_E = \sqrt{\omega_E^2 + \omega_D^2}$ and $cs = cos(k a)$.

After solving the spin-wave eigen equation [Eqs. (\ref{spin-wave eigen-equation for bipartite configuration}), (\ref{combined parameter matrix for bipartite configuration}) and (\ref{M matrix for canted configuration})], all the eigenvalues and corresponding eigenvectors are arranged as $\mathcal{M} \mathcal{V} = \Omega \mathcal{V}$, where $\mathcal{M}$ is the parameter matrix Eq. (\ref{combined parameter matrix for bipartite configuration}) with $\mathcal{M}_{21}$ being Eq. (\ref{M matrix for canted configuration}), $\Omega$ is the diagonal matrix of the eigenvalues, expressed as $\Omega = \operatorname{diag} [ \omega_n^+, - \omega_n^+, \omega_n^-, - \omega_n^-, \omega_p^+, - \omega_p^+, \omega_p^-, - \omega_p^- ]$, and the eigenvector matrix
\begin{widetext}
\begin{equation}
\mathcal{V} = \left( \begin{array}{cccccccc}
\rho_n^+ & \rho_n^+ & - \rho_n^- & - \rho_n^- & - \rho_p^+ & - \rho_p^+ & \rho_p^- & \rho_p^- \\[2.5mm]
- \frac{i}{\omega_n^+} & \frac{i}{\omega_n^+} & \frac{i}{\omega_n^-} & - \frac{i}{\omega_n^-} & - \frac{i}{\omega_p^+} & \frac{i}{\omega_p^+} & \frac{i}{\omega_p^-} & - \frac{i}{\omega_p^-} \\[2.5mm]
- \rho_n^+ & - \rho_n^+ & - \rho_n^- & - \rho_n^- & \rho_p^+ & \rho_p^+ & \rho_p^- & \rho_p^- \\[2.5mm]
i \frac{1}{\omega_n^+} & - i \frac{1}{\omega_n^+} & i \frac{1}{\omega_n^-} & - i \frac{1}{\omega_n^-} & i \frac{1}{\omega_p^+} & - i \frac{1}{\omega_p^+} & i \frac{1}{\omega_p^-} & - i \frac{1}{\omega_p^-} \\[2.5mm]
\rho_n^+ \omega_n^+ & - \rho_n^+ \omega_n^+ & - \rho_n^- \omega_n^- & \rho_n^- \omega_n^- & - \rho_p^+ \omega_p^+ & \rho_p^+ \omega_p^+ & \rho_p^- \omega_p^- & - \rho_p^- \omega_p^- \\[2.5mm]
- i & - i & i & i & - i & - i & i & i \\[2.5mm]
- \rho_n^+ \omega_n^+ & \rho_n^+ \omega_n^+ & - \rho_n^- \omega_n^- & \rho_n^- \omega_n^- & \rho_p^+ \omega_p^+ & - \rho_p^+ \omega_p^+ & \rho_p^- \omega_p^- & - \rho_p^- \omega_p^- \\[2.5mm]
i & i & i & i & i & i & i & i
\end{array} \right).
\end{equation}
\end{widetext}
Here, $\omega_n^\pm$ and $\omega_p^\pm$ [unified in Eq. (\ref{omega of canted configuration})] are the nutational and precessional frequencies, respectively. $\rho_n^\pm$ and $\rho_p^\pm$ are expressed collectively as Eq. (\ref{rho of canted configuration}). The columns of $\mathcal{V}$ correspond sequentially to the eigenvectors associated with the frequencies given in the diagonal matrix of the eigenfrequencies $\Omega$. From $\mathcal{V}$, we then extract the complex amplitudes of SWs, i.e. $\mathcal{A}^e_\theta$, $\mathcal{A}^e_\phi$, $\mathcal{A}^o_\theta$, and $\mathcal{A}^o_\phi$. Substituting these amplitudes into the plane-wave solutions in Eq. (\ref{plane-wave solution}) and taking the real parts yields the explicit expressions for spin-wave excitations, as formulated in Eqs. (\ref{components of spin-wave fluctuation for 2j sublattice})-(\ref{components of spin-wave fluctuation for 2j+1 sublattice}). All relevant parameters are summarized in Tab. \ref{spin-wave parameters of canted configuration}.

 \vspace{-1em}

\section{SW solution in homogeneous-DMI antiferromagnetic configuration} \label{Spin-wave solution in homogeneous-DMI antiferromagnetic configuration}

Compared with Sec. \ref{Spin-wave solution in staggered-DMI antiferromagnetic configuration}, the AFM configuration remains identical, while the DMI term is modified into the staggered form. So, following the same procedure as Sec. \ref{Spin-wave solution in staggered-DMI antiferromagnetic configuration}, the matrixes in Eq. (\ref{Hessian matrix}) can be calculated. $\mathcal{H}_l$ is still $2 \operatorname{diag} [\omega_E + \omega_K, \omega_E + \omega_K]$. $\mathcal{H}_{l \pm 1}$ becomes
\vspace{-0.1em}
\begin{equation}
\mathcal{H}_{l \pm 1} = \left( \begin{array}{cc} \omega_E & \mp \omega_D \cos \phi_l^0 \\ \mp \omega_D \cos \phi_l^0 & - \omega_E \end{array} \right). \label{Hessian-type matrix of homogeneous DMI-AFM configuration}
\end{equation}
\vspace{1em}
Then, the eigen equation for SWs is derived, described by Eqs. (\ref{spin-wave eigen-equation for bipartite configuration}) and (\ref{combined parameter matrix for bipartite configuration}) with
\begin{equation}
\mathcal{M}_{21} = \frac{2}{\eta} \left( \! \begin{array}{cccc}
\omega_E \! + \! \omega_K & 0 & \omega_E cs & i \omega_D sn \\
0 & \omega_E \! + \! \omega_K & i \omega_D sn & - \omega_E cs \\
\omega_E cs & - i \omega_D sn & \omega_E \! + \! \omega_K & 0 \\
- i \omega_D sn & - \omega_E cs & 0 & \omega_E \! + \! \omega_K
\end{array} \! \right), \label{M matrix for homogeneous-DMI antiferromagnetic configuration}
\end{equation}
with $sn = sin(k a)$ and $cs = cos(k a)$. Here, compared with Eq. (\ref{M12 of staggered DMI-AFM configuration}), only the DMI-realted elements are changed.

After solving the spin-wave eigen equation [Eqs. (\ref{spin-wave eigen-equation for bipartite configuration}), (\ref{combined parameter matrix for bipartite configuration}) and (\ref{M matrix for homogeneous-DMI antiferromagnetic configuration})], all the eigenvalues and corresponding eigenvectors are arranged as $\mathcal{M} \mathcal{V} = \Omega \mathcal{V}$, where $\mathcal{M}$ is the parameter matrix Eq. (\ref{combined parameter matrix for bipartite configuration}) with $\mathcal{M}_{21}$ being Eq. (\ref{M matrix for homogeneous-DMI antiferromagnetic configuration}), $\Omega$ is the diagonal matrix of the eigenvalues, expressed as $\Omega = \operatorname{diag} [ \omega_n^+, - \omega_n^+, \omega_n^-, - \omega_n^-, \omega_p^+, - \omega_p^+, \omega_p^-, - \omega_p^- ]$, and the eigenvector matrix is written as
\begin{widetext}
\begin{equation}
\mathcal{V} = \left( \begin{array}{cccccccc}
\frac{\rho_n^+}{\omega_n^+} & - \frac{1}{\rho_n^+ \omega_n^+} & - \frac{\rho_n^-}{\omega_n^-} & \frac{1}{\rho_n^- \omega_n^-} & - \frac{\rho_p^+}{\omega_p^+} & \frac{1}{\rho_p^+ \omega_p^+} & \frac{\rho_p^-}{\omega_p^-} & - \frac{1}{\rho_p^- \omega_p^-} \\[2.5mm]
\frac{i \rho_n^+}{\omega_n^+} & - \frac{i}{\rho_n^+ \omega_n^+} & \frac{i \rho_n^-}{\omega_n^-} & - \frac{i}{\rho_n^- \omega_n^-} & - \frac{i \rho_p^+}{\omega_p^+} & \frac{i}{\rho_p^+ \omega_p^+} & - \frac{i \rho_p^-}{\omega_p^-} & \frac{i}{\rho_p^- \omega_p^-} \\[2.5mm]
\frac{1}{\omega_n^+} & - \frac{1}{\omega_n^+} & - \frac{1}{\omega_n^-} & \frac{1}{\omega_n^-} & \frac{1}{\omega_p^+} & - \frac{1}{\omega_p^+} & - \frac{1}{\omega_p^-} & \frac{1}{\omega_p^-} \\[2.5mm]
- \frac{i}{\omega_n^+} & \frac{i}{\omega_n^+} & - \frac{i}{\omega_n^-} & \frac{i}{\omega_n^-} & - \frac{i}{\omega_p^+} & \frac{i}{\omega_p^+} & - \frac{i}{\omega_p^-} & \frac{i}{\omega_p^-} \\[2.5mm]
\rho_n^+ & \frac{1}{\rho_n^+} & - \rho_n^- & - \frac{1}{\rho_n^-} & - \rho_p^+ & - \frac{1}{\rho_p^+} & \rho_p^- & \frac{1}{\rho_p^-} \\[2.5mm]
i \rho_n^+ & \frac{i}{\rho_n^+} & i \rho_n^- & \frac{i}{\rho_n^-} & - i \rho_p^+ & - \frac{i}{\rho_p^+} & - i \rho_p^- & - \frac{i}{\rho_p^-} \\[2.5mm]
1 & 1 & - 1 & - 1 & 1 & 1 & - 1 & - 1 \\[2.5mm]
- i & - i & - i & - i & - i & - i & - i & - i
\end{array} \right).
\end{equation}
\end{widetext}
Here, $\omega_n^\pm$ and $\omega_p^\pm$, unified in Eq. (\ref{dispersion of homogeneous DMI-AFM configuration}), denote the nutational and precessional frequencies, respectively, while $\rho_n^\pm$ and $\rho_p^\pm$ are collectively given by Eq. (\ref{rho of homogeneous DMI-AFM configuration}). The columns of $\mathcal{V}$ sequentially correspond to the eigenvectors associated with the frequencies listed in the diagonal eigenfrequency matrix $\Omega$. From $\mathcal{V}$, we then extract the complex spin-wave amplitudes, $\mathcal{A}^e_\theta$, $\mathcal{A}^e_\phi$, $\mathcal{A}^o_\theta$, and $\mathcal{A}^o_\phi$. Substituting these amplitudes into the plane-wave solutions in Eq. (\ref{plane-wave solution}) and taking the real parts yields the explicit expressions for the spin-wave excitations, as given in Eqs. (\ref{components of spin-wave fluctuation for 2j sublattice}) and (\ref{components of spin-wave fluctuation for 2j+1 sublattice}). The relevant parameters are summarized in Tab. \ref{spin-wave parameters of homogeneous DMI-AFM configuration}.

\vspace{-1em}

\section{SW solution in spiral configuration} \label{Spin-wave solution in spiral configuration}

When $\omega_D > \omega_D^c$, the static configuration is spiral under homogeneous DMI, as schematically displayed in Fig. \ref{magnetic configurations}(b). The magnetization in equilibrium reads $\mathbf{m}^0_l = \cos (l \phi_s) \mathbf{e}_x + \sin (l \phi_s) \mathbf{e}_y$.

We then combine this equilibrium magnetization with the spin-wave fluctuations in Eq. (\ref{spin-wave fluctuation}) to obtain the total magnetization. Substituting the resulting expression into the reduced magnetic energy for homogeneous DMI in Eq. (\ref{reduced magnetic energy}), together with the coordinate transformation in Eq. (\ref{local frame}), allows all matrices in Eq. (\ref{Hessian matrix}) to be calculated explicitly as follows: $\mathcal{H}_l = - 2 \operatorname{diag} [ \omega_K + \omega_E \cos \phi_s + \omega_D \sin \phi_s, \omega_E \cos \phi_s + \omega_D \sin \phi_s]$, and $\mathcal{H}_{l \pm 1} = \operatorname{diag} [ \omega_E, \omega_E \cos \phi_s + \omega_D \sin \phi_s ]$, where the helical angle $\phi_s$ is Eq. (\ref{spiral angle}).

Next, the parameter matrixes $\mathcal{M}_l$ and $\mathcal{M}_{l \pm 1}$ in Eq. (\ref{spin-wave eigen-equation}) can be constructed using $\mathcal{H}_l$ and $\mathcal{H}_{l \pm 1}$, as well as Eqs. (\ref{symplectic unit matrix}), (\ref{parameter matrix at site l}) and (\ref{parameter matrix at site l +- 1}). It should be noted that $\mathcal{H}_l$ and $\mathcal{H}_{l \pm 1}$ are independent of the site location $l$. Therefore, after calculating the parameter matrixes $\mathcal{M}_l$ and $\mathcal{M}_{l \pm 1}$, Eq. (\ref{spin-wave eigen-equation}) becomes
\begin{equation}
\mathcal{M} \left( \begin{array}{c} \mathcal{A}_\theta \\[0.5em] \mathcal{A}_\phi \\[0.5em] \omega \mathcal{A}_\theta \\[0.5em] \omega \mathcal{A}_\phi \end{array} \right)
= \omega \left( \begin{array}{c} \mathcal{A}_\theta \\[0.5em] \mathcal{A}_\phi \\[0.5em] \omega \mathcal{A}_\theta \\[0.5em] \omega \mathcal{A}_\phi \end{array} \right), \label{spin-wave equation for spiral configuration}
\end{equation}
where the parameter matrix read
\begin{equation}
\mathcal{M} = \frac{1}{\eta} \left( \! \begin{array}{cccc}
0 & 0 & \eta & 0 \\[0.5em]
0 & 0 & 0 & \eta \\[0.5em]
2 \left( \tilde{\omega}_E \! - \! \omega_K \! + \! \omega_E cs \right) & 0 & 0 & i \\[0.5em]
0 & 2 \tilde{\omega}_E \left( 1 \! - \! cs \right) & - i & 0
\end{array} \! \right), \label{M matrix for spiral configuration}
\end{equation}
with $\widetilde{\omega}_E = \sqrt{\omega_E^2 + \omega_D^2}$ and $cs = cos(k a)$.

After solving these linearized equations [Eq. (\ref{spin-wave equation for spiral configuration})], all the eigenvalues and corresponding eigenvectors are arranged as $\mathcal{M} \mathcal{V} = \Omega \mathcal{V}$, where $\Omega$ is the diagonal matrix of the eigenvalues, expressed as $\Omega = \text{diag} [ \omega_n, - \omega_n, \omega_p, - \omega_p ]$, and the eigenvector matrix is given by
\begin{equation}
\mathcal{V} = \left( \begin{array}{cccc}
- \rho_n & - \rho_n & \rho_p & \rho_p \\[2.5mm]
\frac{i}{\omega_n} & - \frac{i}{\omega_n} & \frac{i}{\omega_p} & - \frac{i}{\omega_p} \\[2.5mm]
- \rho_n \omega_n & \rho_n \omega_n & \rho_p \omega_p & - \rho_p \omega_p \\[2.5mm]
i & i & i & i
\end{array} \right).
\end{equation}
Here, $\omega_{n,p}$ are $\omega_{n,p}^-$ in Eq. (\ref{omega of canted configuration}), and $\rho_{n,p}$ are $\rho_{n,p}^-$ in Eq. (\ref{rho of canted configuration}). The columns of $\mathcal{V}$ correspond, in order, to the eigenvectors associated with the frequencies listed in the diagonal eigenfrequency matrix $\Omega$. The site-independent complex spin-wave amplitudes  $\mathcal{A}_\theta$ and $\mathcal{A}_\phi$ are then extracted from $\mathcal{V}$. Substituting these amplitudes into the plane-wave solutions in Eq. (\ref{plane-wave solution}) and taking the real parts yields the explicit spin-wave excitations given in Eq. (\ref{components of spin-wave fluctuation}). The relevant parameters are summarized in Tab. \ref{spin-wave parameters of spiral configuration}.


\begin{thebibliography}{99}

\bibitem{Ritwik Mondal JMMM} Ritwik Mondal, Levente R\'{o}zsa, Michael Farle, Peter M. Oppeneer, Ulrich Nowak, Mikhail Cherkasskii, Inertial effects in ultrafast spin dynamics, J. Magn. Magn. Mater. \textbf{579}, 170830 (2023).

\bibitem{Kumar Neeraj} Kumar Neeraj, Nilesh Awari, Sergey Kovalev, Debanjan Polley, Nanna Zhou Hagstr\"{o}m, Sri Sai Phani Kanth Arekapudi, Anna Semisalova, Kilian Lenz, Bertram Green, Jan-Christoph Deinert, Igor Ilyakov, Min Chen, Mohammed Bawatna, Valentino Scalera, Massimiliano d`Aquino, Claudio Serpico, Olav Hellwig, Jean-Eric Wegrowe, Michael Gensch, and Stefano Bonetti, Inertial spin dynamics in ferromagnets, Nat. Phys. \textbf{17}, 245 (2021).
\bibitem{Vivek Unikandanunni} Vivek Unikandanunni, Rajasekhar Medapalli, Marco Asa, Edoardo Albisetti, Daniela Petti, Riccardo Bertacco, Eric E. Fullerton, and Stefano Bonetti, Inertial spin dynamics in epitaxial cobalt films, Phys. Rev. Lett. \textbf{129}, 237201 (2022).
\bibitem{Anulekha De} Anulekha De, Julius Schlegel, Akira Lentfert, Laura Scheuer, Benjamin Stadtm\"{u}ller, Philipp Pirro, Georg von Freymann, Ulrich Nowak, and Martin Aeschlimann, Magnetic nutation: Transient separation of magnetization from its angular momentum, Phys. Rev. B \textbf{111}, 014432 (2025).

\bibitem{M.-C. Ciornei} M.-C. Ciornei, J. M. Rub\'{i}, and J.-E. Wegrowe, Magnetization dynamics in the inertial regime: Nutation predicted at short time scales, Phys. Rev. B \textbf{83}, 020410(R) (2011).

\bibitem{J.-E. Wegrowe} J.-E. Wegrowe and M.-C. Ciornei, Magnetization dynamics, gyromagnetic relation, and inertial effects, Am. J. Phys. \textbf{80}, 607 (2012).
\bibitem{Stefano Giordano} Stefano Giordano and Pierre-Michel D\'{e}jardin, Derivation of magnetic inertial effects from the classical mechanics of a circular current loop, Phys. Rev. B \textbf{102}, 214406 (2020).

\bibitem{Harry Suhl} Harry Suhl, Theory of the magnetic damping constant, IEEE Trans. Magn. \textbf{34}, 1834 (1998).
\bibitem{Manfred Fahnle} Manfred F\"{a}hnle, Daniel Steiauf, and Christian Illg, Generalized Gilbert equation including inertial damping: Derivation from an extended breathing Fermi surface model, Phys. Rev. B \textbf{84}, 172403 (2011).
\bibitem{Satadeep Bhattacharjee} Satadeep Bhattacharjee, Lars Nordst\"{o}m, and Jonas Fransson, Atomistic Spin Dynamic Method with both Damping and Moment of Inertia Effects Included from First Principles, Phys. Rev. Lett. \textbf{108}, 057204 (2012).
\bibitem{Toru Kikuchi} Toru Kikuchi and Gen Tatara, Spin dynamics with inertia in metallic ferromagnets, Phys. Rev. B 92, 184410 (2015).
\bibitem{Pascal Thibaudeau} Pascal Thibaudeau and Stam Nicolis, Emerging magnetic nutation, Eur. Phys. J. B \textbf{94}, 196 (2021).
\bibitem{J. Anders} J. Anders, C. R. J. Sait, and S. A. R. Horsley, Quantum Brownian motion for magnets, New J. Phys. \textbf{24} 033020 (2022).
\bibitem{Fuming Xu} Fuming Xu, Gaoyang Li, Jian Chen, Zhizhou Yu, Lei Zhang, Baigeng Wang, and Jian Wang, Unified framework of the microscopic Landau-Lifshitz-Gilbert equation and its application to skyrmion dynamics, Phys. Rev. B \textbf{108}, 144409 (2023).
\bibitem{Utkarsh Bajpai} Utkarsh Bajpai and Branislav K. Nikoli\'{c}, Time-retarded damping and magnetic inertia in the Landau-Lifshitz-Gilbert equation self-consistently coupled to electronic time-dependent nonequilibrium Green functions, Phys. Rev. B \textbf{99}, 134409 (2019).
\bibitem{Christian Svingen Johnsen} Christian Svingen Johnsen and Asle Sudb\o, Dynamically generated spin interactions and nutational spin inertia in normal metal-ferromagnet heterostructures, Phys. Rev. B \textbf{111}, 144423 (2025).
\bibitem{Mario Gaspar Quarenta} Mario Gaspar Quarenta, Mithuss Tharmalingam, Tim Ludwig, H. Y. Yuan, Lukasz Karwacki, Robin C. Verstraten, and Rembert A. Duine, Bath-Induced Spin Inertia, Phys. Rev. Lett. \textbf{133}, 136701 (2024).

\bibitem{Ritwik Mondal PRB96} Ritwik Mondal, Marco Berritta, Ashis K. Nandy, and Peter M. Oppeneer, Relativistic theory of magnetic inertia in ultrafast spin dynamics, Phys. Rev. B \textbf{96}, 024425 (2017).
\bibitem{Ritwik Mondal JPCM30} Ritwik Mondal, Marco Berritta, and Peter M Oppeneer, Generalisation of Gilbert damping and magnetic inertia parameter as a series of higher-order relativistic terms, J. Phys.: Condens. Matter \textbf{30}, 265801 (2018).

\bibitem{Felipe Reyes-Osorio} Felipe Reyes-Osorio and Branislav K. Nikoli\'{c}, Optically Induced Magnetic Inertia and Magnons from Non-Markovian Extension of the Landau-Lifshitz-Gilbert Equation, Phys. Rev. Lett. \textbf{135}, 246701 (2025).

\bibitem{S. V. Titov PRB103} S. V. Titov, W. T. Coffey, Yu. P. Kalmykov, and M. Zarifakis, Deterministic inertial dynamics of the magnetization of nanoscale ferromagnets, Phys. Rev. B \textbf{103}, 214444 (2021).
\bibitem{D. Bottcher} D. B\"{o}ttcher, and J. Henk, Significance of nutation in magnetization dynamics of nanostructures, Phys. Rev. B \textbf{86}, 020404(R) (2012).
\bibitem{Sergei V. Titov PRB107} Sergei V. Titov, William J. Dowling, Anton S. Titov, Sergey A. Nikitov, and Mikhail Cherkasskii, Inertial dynamics and equilibrium correlation functions of magnetization at short times, Phys. Rev. B \textbf{107}, 104416 (2023).

\bibitem{E. Olive APL} E. Olive, Y. Lansac, and J.-E. Wegrowe, Beyond ferromagnetic resonance: The inertial regime of the magnetization, Appl. Phys. Lett. \textbf{100}, 192407 (2012).
\bibitem{E. Olive JAP} E. Olive, Y. Lansac, M. Meyer, M. Hayoun, and J.-E. Wegrowe, Deviation from the Landau-Lifshitz-Gilbert equation in the inertial regime of the magnetization, J. Appl. Phys. \textbf{117}, 213904 (2015).
\bibitem{Mikhail Cherkasskii PRB102} Mikhail Cherkasskii, Michael Farle, and Anna Semisalova, Nutation resonance in ferromagnets, Phys. Rev. B \textbf{102}, 184432 (2020).
\bibitem{Ritwik Mondal JPCM33} Ritwik Mondal, Theroy of magnetic inertial dynamics in two-sublattice ferromagnets, J. Phys.: Condens. Matter \textbf{33} 275804 (2021).
\bibitem{Sergei V. Titov JAP} Sergei V. Titov, William J. Dowling, and Yuri P. Kalmykov, Ferromagnetic and nutation resonance frequencies of nanomagnets with various magnetocrystalline anisotropies, J. Appl. Phys. \textbf{131}, 193901 (2022).
\bibitem{Mikhail Cherkasskii PRB106} Mikhail Cherkasskii, Igor Barsukov, Ritwik Mondal, Michael Farle, and Anna Semisalova, Theory of inertial spin dynamics in anisotropic ferromagnets, Phys. Rev. B \textbf{106}, 054428 (2022).
\bibitem{Subhadip Ghosh PRB} Subhadip Ghosh, Mikhail Cherkasskii, Igor Barsukov, and Ritwik Mondal, Theory of tensorial magnetic inertia in terahertz spin dynamics, Phys. Rev. B \textbf{110}, 174430 (2024).
\bibitem{Jonas Wiemeler} Jonas Wiemeler, Michael Farle, and Anna S. Semisalova, High-frequency susceptibility tensor in the inertial regime: Prediction for subterahertz magnonics in ferromagnets, Phys. Rev. B \textbf{112}, 094433 (2025).

\bibitem{Ritwik Mondal PRB103} Ritwik Mondal, Sebastian Gro{\ss}enbach, Levente R\'{o}zsa, and Ulrich Nowak, Nutation in antiferromagnetic resonance, Phys. Rev. B \textbf{103}, 104404 (2021).
\bibitem{Ritwik Mondal PRB10410} Ritwik Mondal and Peter M. Oppeneer, Influence of intersublattice coupling on the terahertz nutation spin dynamics in antiferromagnets, Phys. Rev. B \textbf{104}, 104405 (2021).
\bibitem{David Angster} David Angster, Tobias Dannegger, Julius Schlegel, Martin Evers, and Ulrich Nowak, Nutational resonance modes in antiferromagnetic materials, Sci. Rep. \textbf{15}, 21543 (2025).

\bibitem{Ritwik Mondal PRB10421} Ritwik Mondal, and Akashdeep Kamra, Spin pumping at terahertz nutation resonances, Phys. Rev. B \textbf{104}, 214426 (2021).

\bibitem{Peng-Bin He PRB112} Peng-Bin He, Ri-Xing Wang, Zai-Dong Li, and Mikhail Cherkasskii, Chirality and polarization of inertial antiferromagnetic resonances driven by spin-orbit torques, Phys. Rev. B \textbf{112}, 224423 (2025).

\bibitem{Peng-Bin He PRB108} Peng-Bin He, Large-amplitude and widely tunable self-oscillations enabled by the inertial effect in uniaxial antiferromagnets driven by spin-orbit torques, Phys. Rev. B \textbf{108}, 184418 (2023).
\bibitem{Peng-Bin He PRB11006} Peng-Bin He, Influence of the magnetic inertia on the self-oscillation in spin-orbit torque-driven tripartite antiferromagnets with a $120^\circ$ rotation symmetry, Phys. Rev. B \textbf{110}, 064411 (2024).
\bibitem{Rodolfo Rodriguez} Rodolfo Rodriguez, Mikhail Cherkasskii, Rundong Jiang, Ritwik Mondal, Arezoo Etesamirad, Allison Tossounian, Boris A. Ivanov, and Igor Barsukov, Spin Inertia and Auto-Oscillations in Ferromagnets, Phys. Rev. Lett. \textbf{132}, 246701 (2024).

\bibitem{Rahnuma Rahman} Rahnuma Rahman and Supriyo Bandyopadhyay, An observable effect of spin inertia in slow magneto-dynamics: increase of the switching error rates in nanoscale ferromagnets, J. Phys.: Condens. Matter \textbf{33} 355801 (2021).
\bibitem{Kumar Neeraj PRB} Kumar Neeraj, Matteo Pancaldi, Valentino Scalera, Salvatore Perna, Massimiliano d'Aquino, Claudio Serpico, and Stefano Bonetti, Magnetization switching in the inertial regime, Phys. Rev. B \textbf{105}, 054415 (2022).
\bibitem{I. Makhfudz PRB} I. Makhfudz, Y. Hajati, and E. Olive, High-temperature magnetization reversal in the inertial regime, Phys. Rev. B \textbf{106}, 134415 (2022).
\bibitem{Lucas Winter} Lucas Winter, Sebastian Gro{\ss}enbach, Ulrich Nowak, and Levente R\'{o}zsa, Nutational switching in ferromagnets and antiferromagnets, Phys. Rev. B \textbf{106}, 214403 (2022).

\bibitem{I. Makhfudz APL} I. Makhfudz, E. Olive, and S. Nicolis, Nutation wave as a platform for ultrafast spin dynamics in ferromagnets, Appl. Phys. Lett. \textbf{117}, 132403 (2020).
\bibitem{Mikhail Cherkasskii PRB103} Mikhail Cherkasskii, Michael Farle, and Anna Semisalova, Dispersion relation of nutation surface spin waves in ferromagnets, Phys. Rev. B \textbf{103}, 174435 (2021).
\bibitem{Alexey M. Lomonosov} Alexey M. Lomonosov, Vasily V. Temnov, and Jean-Eric Wegrowe, Anatomy of inertial magnons in ferromagnetic nanostructures, Phys. Rev. B \textbf{104}, 054425 (2021).
\bibitem{Sergei V. Titov PRB105} Sergei V. Titov, William J. Dowling, Yuri P. Kalmykov, and Mikhail Cherkasskii, Nutation spin waves in ferromagnets, Phys. Rev. B \textbf{105}, 214414 (2022).
\bibitem{Ritwik Mondal PRB106} Ritwik Mondal, and Levente R\'{o}zsa, Inertial spin waves in ferromagnets and antiferromagnets, Phys. Rev. B \textbf{106}, 134422 (2022).
  \bibitem{Mikhail Cherkasskii PRB109} Mikhail Cherkasskii, Ritwik Mondal, and Levente R\'{o}zsa, Inertial spin waves in spin spirals, Phys. Rev. B \textbf{109}, 184424 (2024).
\bibitem{Peng-Bin He PRB110} Peng-Bin He and Mikhail Cherkasskii, Temporal and spatial attenuation of inertial spin waves driven by spin-transfer torques, Phys. Rev. B \textbf{110}, 174431 (2024).
\bibitem{Massimiliano d'Aquino} Massimiliano d'Aquino, and Riccardo Hertel, Nonreciprocal Inertial Spin-Wave Dynamics in Twisted Magnetic Nanostrips, Phys. Rev. Lett. \textbf{135}, 216705 (2025).
\bibitem{H. Y. Yuan} H. Y. Yuan, Using surface plasmons to detect spin inertia, Phys. Rev. B \textbf{112}, 054438 (2025).
\bibitem{Subhadip Ghosh JPCM} Subhadip Ghosh, Darpa Narayan Basu, and Ritwik Mondal, Engineering spin-wave spectrum via the magnetization inertia tensor, J. Phys.: Condens. Matter \textbf{38}, 015803 (2026).

\bibitem{D. H. Goldstein} D. H. Goldstein, \textit{Polarized light}, (CRC Press, Taylor \text{\&} Francis Group, 2011), p. 57.

\bibitem{I. Dzyaloshinskii} I. Dzyaloshinskii, A thermodynamic theory of ``weak" ferromagnetism of antiferromagnetics, J. Phys. Chem. Solids \textbf{4}, 241 (1958).
\bibitem{Toru Moriya} T\^{o}ru Moriya, Anisotropic Superexchange Interaction and Weak Ferromagnetism, Phys. Rev. \textbf{120}, 91 (1960).
\bibitem{J. H. H. Perk} J. H. H. Perk and H. W. Capel, Antisymmetric exchange, canting and spiral structure, Phys. Lett. A \textbf{58}, 115 (1976).
\bibitem{Carmine Autieri} Carmine Autieri, Raghottam M. Sattigeri, Giuseppe Cuono, and Amar Fakhredine, Staggered Dzyaloshinskii-Moriya interaction inducing weak ferromagnetism in centrosymmetric altermagnets and weak ferrimagnetism in noncentrosymmetric altermagnets, Phys. Rev. B \textbf{111}, 054442 (2025).
\bibitem{Xiyin Ye} Xiyin Ye, Qirui Cui, Weiwei Lin, and Tao Yu, Spin quenching and transport by hidden Dzyaloshinskii-Moriya interactions, Phys. Rev. B \textbf{111}, 064401 (2025).

\bibitem{arctan} Wolfram Research (1988), ArcTan, Wolfram Language function, \url{https://reference.wolfram.com/language/ref/ArcTan.html} (updated 2021).

\bibitem{Sergio M. Rezende} Sergio M. Rezende, Antonio Azevedo, and Roberto L. Rodr\'{i}guez-Su\'{a}rez, Introduction to antiferromagnetic magnons, J. Appl. Phys. \textbf{126}, 151101 (2019).

\bibitem{Priyanka Vaidya} Priyanka Vaidya, Sophie A. Morley, Johan van Tol, Yan Liu, Ran Cheng, Arne Brataas, David Lederman, and Enrique del Barco, Subterahertz spin pumping from an insulating antiferromagnet, Science \textbf{368}, 160 (2020).

\bibitem{Z. Q. Qiu} Z. Q. Qiu, Chirality dependence of spin current in spin pumping, Nat. Commun. \textbf{13}, 5229 (2022).

\bibitem{Yutian Wang} Yutian Wang, Jiongjie Wang, Ruoban Ma, and Jiang Xiao, Connection between spin-wave polarization and dissipation, Phys. Rev. B \textbf{111}, 134431 (2025).

\bibitem{Taku J Sato} Taku J Sato and Kittiwit Matan, Nonreciprocal Magnons in Noncentrosymmetric Magnets, J. Phys. Soc. Jpn. \textbf{88}, 081007 (2019).

\bibitem{J. J. Sakurai} J. J. Sakurai, \textit{Modern Quantum Mechanics}, Revised edition (Addison-Wesley, 1994), pp. 30-41.

\bibitem{Roald Hoffmann} Roald Hoffmann, \textit{Solids and Surfaces: A Chemist's View of Bonding in Extended Structures} (Wiley-VCH, 1988), pp. 83-90.

\bibitem{R. W. Damon} R. W. Damon and J. R. Eshbach, J. Phys. Chem. Solids \textbf{19}, 308 (1961).








\end{thebibliography}
\end{document}